\documentclass[useAMS,usenatbib]{mn2e}

\usepackage{epsfig,graphicx,graphics}

\title[The 31 and 250~GHz continua in PNe]{A
  centimetre-wave excess over free-free emission in planetary nebulae}
\author[S. Casassus et al.]{
S. Casassus$^{1}$\thanks{E-mail: simon@das.uchile.cl (SC)},
L.-{\AA}. Nyman$^{2,3}$, C. Dickinson$^{4,5}$, T. J.  Pearson$^4$
\\
$^{1}$ Departamento de Astronom\'{\i}a, Universidad de Chile, Casilla 36-D, Santiago, Chile\\ 
$^{2}$ European Southern Observatory, Casilla 19001, Santiago 19, Chile\\
$^{3}$ Onsala Space Observatory, S-439 92 Onsala, Sweden\\
$^{4}$ Chajnantor Observatory, M/S 105-24, California Institute of
  Technology, Pasadena, CA 91125\\
$^{5}$ Jet Propulsion Laboratory, M/S 169-327, 4800 Oak Grove Drive,  Pasadena, CA 91109
}

\begin{document}

\date{}

\pagerange{\pageref{firstpage}--\pageref{lastpage}} \pubyear{2006}

\maketitle

\label{firstpage}

\begin{abstract}

We report a centimetre-wave (cm-wave, 5--31~GHz) excess over free-free
emission in PNe.  Accurate 31 and 250~GHz measurements show that the
31~GHz flux densities in our sample are systematically higher than the
level of optically thin free-free continuum extrapolated from
250~GHz. The 31~GHz excess is observed, within one standard deviation,
in all 18 PNe with reliable 31 and 250~GHz data, and is significant in
9 PNe. The only exception is the peculiar object M~2-9, whose radio
spectrum is that of an optically thick stellar wind. On average the
fraction of non-free-free emission represents $51$\% of the total flux
density at 31~GHz, with a scatter of 11\%. The average 31--250~GHz
spectral index of our sample is $\langle \alpha_{31}^{250} \rangle =
-0.43 \pm 0.03$ (in flux density, with a scatter of 0.14).  The
31--250~GHz drop is reminiscent of the anomalous foreground observed
in the diffuse ISM by CMB anisotropy experiments. The 5--31~GHz
spectral indices are consistent with both flat spectra and spinning
dust emissivities, given the 10\% calibration uncertainty of the
comparison 5~GHz data.  But a detailed study of the objects with the
largest cm-excess, including the low frequency data available in the
literature, shows that present spinning dust models cannot alone
explain the cm-excess in PNe.  Although we have no definitive
interpretation of our data, the least implausible explanation involves
a synchrotron component absorbed by a cold nebular screen.  We give
flux densities for 37 objects at 31~GHz, and for 26 objects at
250~GHz.

\end{abstract}


\section{Introduction}

An increasing amount of evidence supports the existence of a new
continuum emission mechanism in the diffuse interstellar medium (ISM)
at 10--30~GHz, other than free-free, synchrotron, or an hypothetical
Rayleigh-Jeans tail of cold dust grains \citep[e.g.,][]{lei97,fin04,
  wat05}. As proposed by \citet{dl98b} a promising candidate emission
mechanism is electric dipole radiation from spinning very small grains
(VSGs), or spinning dust. Cosmic Background Imager (CBI) observations
of the dark cloud LDN~1622 linked, on morphological grounds, the
cm-wave emitters to the VSGs \citep{cas06}.

The SED of the Helix nebula (NGC~7293, an evolved planetary nebula,
PN) also exhibits a 31~GHz excess over the free-free continuum
expected from 250~GHz measurements \citep{cas04}.  Resolved 31~GHz
images of the Helix, obtained with the CBI, provided morphological
evidence for a new radio continuum component.

PNe have traditionally served as test laboratories for nebular
astrophysics. Their spectra are the product of the photoionisation by
a single exciting star of its surrounding point-symmetric nebula, in a
characteristic radial and homologous expansion.  PNe are archetypical
free-free sources; the interpretation of their $<14~$GHz continuum in
terms of free-free emission is paradigmatic
\citep[e.g.][]{sio01,con98}. Yet inconsistencies with the free-free
paradigm exist in the literature, although not recognised as
such. \citet{hoa92}, in an investigation of the cm- to mm-wave SED of
PNe, ascribed deviations from free-free spectra to instrumental
uncertainties. \citet{gar89} compared continuum and
radio-recombination line maps, finding systematic differences
interpreted as non-LTE amplification or large $T_e$ variations.

Here we report on a survey of the cm- and mm-wave continuum in PNe
acquired with the Cosmic Background Imager \citep[CBI,][]{pad02} and
with the Sest IMaging Bolometer Array \citep[SIMBA,][]{nym01}, at the
Swedish-ESO Submillimetre Telescope (SEST). The PN sample is mostly
selected from \citet{cas01}, and represents the population of compact
and mid-IR-bright PNe (it is not a complete sample).
Section~\ref{sec:index} discusses PN electron temperatures and
relevant free-free spectral indices. Section~\ref{sec:cbi} and
\ref{sec:simba} present the CBI and SIMBA data,
respectively. Section~\ref{sec:comp} compares the two datasets, which
are in turn compared to previous measurements in
Section~\ref{sec:previous}. Section~\ref{sec:candidates} interprets
our findings in terms of possible emission
mechanisms. Section~\ref{sec:indiv} analyses the SEDs of individual
PNe. Section~\ref{sec:conc} concludes. All the data analysis was
carried out using Perl Data Language ({\tt http://www.pdl.org}),
unless otherwise stated. The error bars on all Figures are one
standard deviation ($\pm$1~$\sigma$, the error bars have a total
length of 2~$\sigma$).

\section{Adopted planetary nebula electron temperatures and free-free spectral
  indices} \label{sec:index}

The radio-frequency free-free spectral index is roughly constant at
$\alpha \sim -0.15$ in flux density, $F(\nu) = F(\nu_\circ) (\nu /
\nu_\circ)^\alpha$. Yet $\alpha$ has a residual dependence on
frequency, electron temperature $T_e$, and Helium contribution.  Since
our aim is to perform a precision analysis of the PN SEDs, we
calculated accurate free-free indices for the frequencies relevant to
this work. The values we obtained are exact for non-relativistic
electrons (i.e. $T_e < 10^8$~K).  Full details are given in
Sec.~\ref{sec:exactindex}.

The electron temperature in PNe has been the matter of significant
debate.  \citet{pei71} first observed that optical recombination lines
(ORLs) and collisionally-excited lines (CELs) give broadly different
$T_e$ values. Since then, several new $T_e$ diagnostics have been
reported, all pointing towards the conclusion that ORLs and continuum
diagnostics give systematically lower $T_e$ values than [O\,{\sc iii}]
$\lambda 4363/\lambda 4959$, by a factor of $\sim 2$ (and up to 10).

The current view, as summarised by \citet{liu06}, is that of a
bi-abundance model in which PNe (and to a lesser extent H\,{\sc ii}
regions) contain dense, hydrogen-deficient (with metalicity $Z \sim
1$) and cool inclusions accounting for most of the ORL emission. In
the bi-abundance model the bulk of the nebula, by mass, is best
represented by the CEL diagnostics, with $T_e \sim 10000-15000~$K. In
the case of NGC~6153, the free-free luminosity of the ORL emitting gas
is $\sim$1/10 that of the CEL emitting gas, and the hydrogen mass in
the ORL gas is only $\sim 1-2~$\% that of the CEL gas \citep[Mike
Barlow, private communication, and][ their model IH3]{liu00}.
Assuming the ORL $T_e$ is $\sim$1/2 the CEL $T_e$, an average $T_e$,
weighted by free-free luminosity gives, in the case of NGC~6153,
$\langle T_e \rangle = 0.95 T_e(\mathrm{CEL})$.

Our strategy was to adopt a single conservative value for the whole PN
sample. We thus adopted $T_e = 7000~$K, a value that approaches a
canonical $T_e$ for H\,{\sc ii} regions \citep[e.g.,][]{dic03}, but is
unrealistically cold for PNe. This choice minimises free-free
deviations in the dataset, as cooler $T_e$ give steeper 31-250~GHz
spectral indices.  Since the emissivity of the He$^+$ continuum gives
somewhat steeper indices than for H, we also included a contribution
from the He$^+$ continuum, assuming a large but conservative 10\%
abundance of He$^{++}$ relative to H$^+$.  Table~\ref{table:indices}
summarises the free-free spectral indices relevant to our analysis
(see Sec.~\ref{sec:finalindices} for details).

\begin{table}
\centering
\caption{Observed and free-free spectral indices at $T_e =
  7000~$K. }
\label{table:indices}
\begin{tabular}{ccc}
\hline 
$\alpha_{5}^{31}$   & $\alpha_{26}^{36}$ & $\alpha_{31}^{250}$ \\ \hline
\multicolumn{3}{c}{Free-free indices} \\
$-0.125^a$  & $-0.137$ & $-0.152$ \\ 
\multicolumn{3}{c}{PN sample} \\
$-0.06\pm0.01(0.22)^{b,c}$ &  $-0.18\pm0.03(0.21)^d$ & $ -0.43\pm0.03(0.14)^e $ \\ \hline
 \end{tabular}
\begin{flushleft}
$^a$ The free-free indices bear a 1$\sigma$ (one standard
  deviation) uncertainty of $10^{-3}$, stemming from the uncertain
  contribution of the He$^+$ free-free continuum.\\
$^b$ The root-mean-square (rms) scatter is indicated in parenthesis.\\
$^c$ From all 31 objects with 5~GHz flux densities and
  $\chi^2_\mathrm{CBI} < 2$. \\
$^d$ From all 32 objects with $\chi^2_\mathrm{CBI} < 2$.\\
$^e$ From all  17 objects with 250~GHz flux densities and
  $\chi^2_\mathrm{CBI} < 2$. \\
\end{flushleft}
\end{table}
\normalsize

\section{CBI observations} \label{sec:cbi}

The CBI is a planar interferometer array with 13 antennas, each 0.9~m
in diameter, mounted on a 6~m tracking platform, which rotates in
parallactic angle to provide uniform $uv$-coverage. Its uniform-weight
synthesised beam is $\sim$6~arcmin. The CBI receivers operate in 10
frequency channels, with 1~GHz bandwidth each, giving a total
bandwidth of 26--36~GHz.  It is located in Llano de Chajnantor,
Atacama, Chile. In this Section we describe the analysis of total
intensity and polarisation observations of PNe.

\subsection{Total intensity}

We acquired CBI total intensity observations of a sample of 36 PNe,
whose flux densities are listed in Table~\ref{table:master}. The
visibility data were reduced and edited using a special-purpose
package (CBICAL, developed by T.J. Pearson).  Flux calibration was
obtained by reference to Mars or Jupiter, and tied to the {\em WMAP}
temperature for Jupiter, $T_\mathrm{Jup} = 147.3 \pm 1.8~$K
\citep[][and references therein]{rea04}. The fractional uncertainty on
$T_\mathrm{Jup}$, $\sigma(T_\mathrm{Jup}) = 1.2~\%$, affects the
overall calibration of the CBI.

Approximate cancellation of ground and Moon contamination was obtained
by differencing with a reference field at the same declination but
offset in hour angle by the duration of the on-source integration. In
general we used an on-source integration time of 8~min. We found that
for point sources the acquisition of a reference field is not
mandatory, since restricting to the longer baselines avoids any ground
contamination \citep[e.g.,][]{cas06}. We thus used the differenced
dataset when available, and discarded baselines shorter than
200~$\lambda$ otherwise. The baseline-restricted and differenced
datasets have similar signal-to-noise ratios on point source
measurements. 

We extracted integrated flux densities by fitting parametrised models
to the visibilities. The uncertainties on the inferred flux densities
correspond to the $\Delta \chi^2 = 1$ contour. Point source models gave
better reduced $\chi^2$ values than elliptical gaussians or core-halo
models, except for NGC~1360, which we fit with a
$4.9\times2.3$~arcmin$^2$ elliptical Gaussian (full width half
maximum, FWHM), with its major axis at a position angle of 22.9~deg,
positive East of North.

The use of point-source models to extract flux densities assumes there
is no contaminating emission within the CBI primary beam, which has a
FWHM of 45.2~arcmin. Additional sources within the primary beam result
in high reduced $\chi^2$ values ($\chi^2_\mathrm{CBI}$ hereafter). We
checked by inspection that extended emission is present for
$\chi^2_\mathrm{CBI}$ greater than $\sim$2.0, which is the cutoff we
adopted to select reliable point source flux densities.  We attempted
to extract flux densities from images of the fields with extended
emission, and obtained the same value as listed in
Table~\ref{table:master}. The extended emission fills the CBI primary
beam, is featureless and difficult to separate from the PN. In
addition to the extended emission, there are also bright point sources
near NGC~6537. 19W32 is altogether drowned by the diffuse
background. Thus the sky-plane measurements have the same degree of
unreliability as the visibility fits.


A measurement is considered reliable if its uncertainty, as estimated
from the point-source models, indeed corresponds to a 1~$\sigma$
deviation. Several integrations on different nights for a subsample of
22 objects confirmed that, when $\chi^2_\mathrm{CBI} < 2.0$, all
nightly measurements are consistent within 3~$\sigma$ with the flux
densities derived from model-fitting the combined visibility dataset.
We also checked for consistency between the differenced and
baseline-restricted datasets.

In addition to the statistical uncertainties due to thermal noise, the
final error bars also include uncertainties in the overall calibration
of the CBI for the PNe observations. The systematics stem in part from
uncertainties in our primary calibrator (Jupiter), and in part from
gain variations during a given night. We thus compared PN flux
densities obtained with two different calibrators, during the same
night.  We measured the fractional flux density differences,
$\Delta_\mathrm{gain} = | F_\mathrm{cal1} - F_\mathrm{cal2} | /
F_\mathrm{cal1}$, when calibrating against calibrator 1 or 2.  On the
night of 15-Jan-2003, NGC~3918 was observed along with Jupiter and
Tau~A; on 3-Jun-2001 NGC3242 was observed along with J1230+123 and
Mars. Using the same reduction (CBICAL) scripts and analysis tools as
those used to produce the flux density reported in
Table~\ref{table:master}, we obtain $\Delta_\mathrm{gain} = 0.68\%$
for NGC~3918, and 0.80\% for NGC~3242. For the PNe observations, the
uncertainty on the CBI response is thus of order $\sigma_\mathrm{gain}
= 1\%$.

Including the uncertainty on $T_\mathrm{Jup}$, we take as a
conservative estimate of fractional systematic uncertainties
$\sigma_\mathrm{systematic} = 2\%$. The final uncertainty on the flux
densities $F_\mathrm{CBI}$ reported in Table~\ref{table:master} is
$\sigma_\mathrm{total}^2 =
(\sigma_\mathrm{systematic}\,F_\mathrm{CBI})^2 + \sigma^2$, where
$\sigma$ derives from model-fitting the combined visibility datasets.

The model-fit spectral indices over 26--36~GHz are all consistent with
free-free emission (see Table~\ref{table:indices}).  The accuracy of
the CBI indices reaches $\sim$0.1 for NGC~6369, NGC~6572, NGC~7009,
IC~418, and NGC~3242.

The CBI~31~GHz and comparison 5~GHz flux densities are also consistent
with optically thin free-free, within the 10\% calibration uncertainty
of the literature data (Table~\ref{table:indices}).  The only
exceptions are Pe~1-7, M~2-9, SwSt~1, and Vy~2-2, all with excesses at
31~GHz and point-source models with $\chi^2_\mathrm{CBI} < 2$.

In a comparison with the 14~GHz dataset, we obtain $\langle
\alpha_5^{14} \rangle = 0.00 \pm 0.03$ with an rms scatter of 0.19,
and $\langle \alpha_{14}^{31} \rangle = -0.23 \pm 0.03$ with a scatter
of 0.37. $\langle \alpha_{14}^{31} \rangle$ is steeper than the
free-free value by 3.3~$\sigma$, if the uncertainties on the 14~GHz
are indeed smaller than 10\% rms.

\begin{table*}
  \small
 \caption{Summary of the CBI and SIMBA measurements. Reduced $\chi^2$
   values, $\chi^2_\mathrm{CBI}$, refer to parametrised models of the
   CBI visibilities. All flux densities are reported in mJy.}
 \label{table:master}
 \begin{tabular}{llccrrcrrrrr}
 \hline
 name & PNG    & 5$^a$ & 14$^a$ & 31~GHz  & $\alpha_{26}^{36}$ &
 $\chi^2_\mathrm{CBI}$ & 250~GHz & $\Delta^b$ & $\Delta/\sigma(\Delta)$  & $\delta^c$  & $\alpha_{31}^{250}$   \\ 
  \hline
       SwSt1 & $   001.5-06.7   $ &      130  &      240  & $    190\pm   6 $  & $ -0.33\pm0.30  $&  1.18    &   $ 14 \pm 102  $   & $   -125\pm 102   $   &  $ ( -1.2 ) $ &$ 90\pm73 $ & $ -1.3\pm 3.5 $    \\  
   NGC6369 & $   002.4+05.8   $ &     2002  &     1718  & $   1452\pm  30 $  & $ -0.20\pm0.06  $&  1.20    &   $ 479 \pm 54  $   & $   -583\pm  58   $   &  $ ({\bf -10.0 }) $ &$ 55\pm 5 $ & $ -0.5\pm 0.1 $    \\  
       Hb4 & $   003.1+02.9   $ &      170  &      148  & $    138\pm   5 $  & $ 0.07\pm0.39  $&  1.11    \\ 
       Hb6 & $   007.2+01.8   $ &      243  &      241  & $    195\pm   6 $  & $ -0.40\pm0.25  $&  1.20    \\ 
   NGC6309 & $   009.6+14.8   $ &      102  &      146  & $     95\pm   7 $  & $ -1.37\pm0.87  $&  1.08    &   $ 16 \pm 16  $   & $    -54\pm  17   $   &  $ ({\bf -3.2 }) $ &$ 77\pm23 $ & $ -0.9\pm 0.5 $    \\  
      M2-9 & $   010.8+18.0   $ &       36  &       44  & $     55\pm   4 $  & $ 0.05\pm0.84  $&  1.11    &   $ 273 \pm 48  $   & $    233\pm  48   $   &  $ ({\bf 4.8 }) $ &$ -582\pm129 $ & $  0.8\pm 0.1 $    \\  
   NGC6537 & $   010.1+00.7   $ &      610  &      557  & $    477\pm  12 $  & $ -0.63\pm0.15  $&  2.13    &   $ 331 \pm 111  $   & $    -17\pm 111   $   &  $ ( -0.2 ) $ &$  5\pm32 $ & $ -0.2\pm 0.2 $    \\  
   NGC6818 & $   025.8-17.9   $ &      281  &      270  & $    270\pm  18 $  & $ -1.04\pm0.69  $&  1.47    &            $<$36  &   &  &    \\  
   NGC6741 & $   033.8-02.6   $ &      220  &      183  & $    181\pm  13 $  & $ -1.42\pm0.79  $&  1.21    \\ 
   NGC6572 & $   034.6+11.8   $ &     1260  &        -  & $   1073\pm  22 $  & $ -0.10\pm0.06  $&  1.16    &   $ 442 \pm 54  $   & $   -343\pm  56   $   &  $ ({\bf -6.1 }) $ &$ 44\pm 7 $ & $ -0.4\pm 0.1 $    \\  
   NGC6790 & $   037.8-06.3   $ &      240  &      256  & $    267\pm   8 $  & $ 0.05\pm0.24  $&  1.20    &   $ 118 \pm 60  $   & $    -77\pm  60   $   &  $ ( -1.3 ) $ &$ 40\pm31 $ & $ -0.4\pm 0.2 $    \\  
   NGC7009 & $   037.7-34.5   $ &      750  &      649  & $    512\pm  11 $  & $ -0.23\pm0.08  $&  1.26    &   $ 157 \pm 27  $   & $   -217\pm  28   $   &  $ ({\bf -7.7 }) $ &$ 58\pm 7 $ & $ -0.6\pm 0.1 $    \\  
     CN3-1 & $   038.2+12.0   $ &       65  &       62  & $     55\pm   5 $  & $ -0.39\pm0.99  $&  1.15    \\ 
     Vy2-2 & $   045.4-02.7   $ &       50  &        -  & $    234\pm   6 $  & $ -0.20\pm0.21  $&  1.18    &   $ 144 \pm 104  $   & $    -27\pm 104   $   &  $ ( -0.3 ) $ &$ 16\pm61 $ & $ -0.2\pm 0.3 $    \\  
     M1-71 & $   055.5-00.5   $ &      204  &        -  & $    178\pm   7 $  & $ 0.37\pm0.35  $&  1.17    \\ 
    IC4997 & $   058.3-10.9   $ &      100  &      127  & $     81\pm   4 $  & $ -0.20\pm0.49  $&  1.14    &   $ 64 \pm 64  $   & $      4\pm  64   $   &  $ ( 0.1 ) $ &$ -7\pm108 $ & $ -0.1\pm 0.5 $    \\  
   NGC6886 & $   060.1-07.7   $ &      105  &      102  & $     58\pm   5 $  & $ -0.82\pm1.00  $&  1.15    \\ 
    NGC246 & $   118.8-74.7   $ &      247  &      248  & $    104\pm  12 $  & $ -1.20\pm1.30  $&  0.32    \\ 
     IC418 & $   215.2-24.2   $ &     1613  &     1529  & $   1334\pm  30 $  & $ -0.39\pm0.11  $&  1.22    &   $ 704 \pm 83  $   & $   -271\pm  86   $   &  $ ({\bf -3.2 }) $ &$ 28\pm 9 $ & $ -0.3\pm 0.1 $    \\  
   NGC1360 & $   220.3-53.9   $ &        -  &        -  & $    132\pm  16 $  & $ 1.16\pm1.35  $&  1.19    \\ 
    IC2165 & $   221.3-12.3   $ &      188  &      186  & $    147\pm   7 $  & $ 0.18\pm0.53  $&  1.12    \\ 
   NGC2440 & $   234.8+02.4   $ &      370  &      325  & $    267\pm  13 $  & $ 0.50\pm0.47  $&  1.29    &   $ 251 \pm 64  $   & $     56\pm  65   $   &  $ ( 0.9 ) $ &$ -29\pm33 $ & $ -0.0\pm 0.1 $    \\  
   NGC3242 & $   261.0+32.0   $ &      896  &      739  & $    544\pm  12 $  & $ -0.33\pm0.09  $&  1.25    &   $ 212 \pm 34  $   & $   -186\pm  35   $   &  $ ({\bf -5.3 }) $ &$ 47\pm 9 $ & $ -0.5\pm 0.1 $    \\  
   NGC3132 & $   272.1+12.3   $ &      235  &      198  & $    184\pm  20 $  & $ 1.81\pm1.19  $&  1.34    &   $ 161 \pm 66  $   & $     27\pm  68   $   &  $ ( 0.4 ) $ &$ -20\pm51 $ & $ -0.1\pm 0.2 $    \\  
   NGC2867 & $   278.1-05.9   $ &      252  &      265  & $    194\pm  20 $  & $ 0.73\pm1.08  $&  1.31    &   $ 21 \pm 56  $   & $   -121\pm  58   $   &  $ ( -2.1 ) $ &$ 85\pm39 $ & $ -1.1\pm 1.3 $    \\  
    IC2501 & $   281.0-05.6   $ &      261  &      236  & $    167\pm   6 $  & $ -0.03\pm0.33  $&  1.11    &   $ 69 \pm 29  $   & $    -53\pm  29   $   &  $ ( -1.8 ) $ &$ 43\pm24 $ & $ -0.4\pm 0.2 $    \\  
   Hen2-47 & $   285.6-02.7   $ &      170  &      187  & $    136\pm   9 $  & $ 0.59\pm0.73  $&  1.17    \\ 
    IC2621 & $   291.6-04.8   $ &      195  &      175  & $    165\pm   5 $  & $ 0.23\pm0.27  $&  1.10    &   $ 22 \pm 30  $   & $    -99\pm  30   $   &  $ ({\bf -3.3 }) $ &$ 82\pm25 $ & $ -1.0\pm 0.7 $    \\  
   NGC3918 & $   294.6+04.7   $ &      859  &      765  & $    626\pm  19 $  & $ -0.44\pm0.25  $&  1.36    &   $ 167 \pm 40  $   & $   -291\pm  42   $   &  $ ({\bf -6.9 }) $ &$ 63\pm 9 $ & $ -0.6\pm 0.1 $    \\  
   NGC5315 & $   309.1-04.3   $ &      480  &      366  & $    443\pm  22 $  & $ 0.51\pm0.52  $&  1.24    &   $ 141 \pm 38  $   & $   -183\pm  41   $   &  $ ({\bf -4.4 }) $ &$ 56\pm12 $ & $ -0.5\pm 0.1 $    \\  
  Hen2-113 & $   321.0+03.9   $ &      115  &      160  & $    133\pm   5 $  & $ 0.05\pm0.34  $&  1.15    \\ 
  Hen2-142 & $   327.1-02.2   $ &       65  &       68  & $     65\pm   4 $  & $ 2.12\pm0.70  $&  1.76    \\ 
     Pe1-7 & $   337.4+01.6   $ &      117  &      119  & $    111\pm   6 $  & $ 0.70\pm0.54  $&  1.36    \\ 
   NGC6153 & $   341.8+05.4   $ &      477  &      559  & $    475\pm  20 $  & $ 0.71\pm0.43  $&  2.40    &   $ 208 \pm 38  $   & $   -139\pm  41   $   &  $ ({\bf -3.4 }) $ &$ 40\pm11 $ & $ -0.4\pm 0.1 $    \\  
   NGC6072 & $   342.1+10.8   $ &        -  &        -  & $     64\pm  34 $  & $ -1.03\pm5.12  $&  1.74    &            $<$87  &   &  &    \\  
   NGC6302 & $   349.5+01.0   $ &     3100  &     3034  & $   2578\pm  53 $  & $ 0.19\pm0.05  $&  2.98    &   $ 1963 \pm 199  $   & $     78\pm 203   $   &  $ ( 0.4 ) $ &$ -4\pm11 $ & $ -0.1\pm 0.0 $    \\  
     H1-12 & $   352.6+00.1   $ &      719  &        -  &   - & - & -   &   $ 383 \pm 65  $   & $    -17\pm  76 $  & &      \\  
     M1-26 & $   358.9-00.7   $ &      400  &      420  &   - & - & -   &   $ 346 \pm 81  $   & $    124\pm  84 $  & &      \\  
     19W32 & $   359.2+01.2   $ &       21  &        -  & $     18\pm   3 $  & $ -10.27\pm1.13  $&  4.34    \\ 
       Hb5 & $   359.3-00.9   $ &      548  &      551  &   - & - & -   &   $ 506 \pm 71  $   & $    201\pm  77 $  & &      \\  

  \hline
 \end{tabular}

\medskip

\begin{flushleft}
$^a$ 5~GHz and 14~GHz measurements from the compilation of
  \cite{ack92}, taken mostly from the Parkes~64m data of
  \citet{mil82}, and bearing a $\sim$10\% calibration uncertainty.

$^b$$\Delta$ is the difference between the observed 250~GHz flux
  density and optically thin free-free emission extrapolated from
  31~GHz, if available, or 5~GHz otherwise.

$^c$ $\delta = 100 \times \Delta / (F(250~\mathrm{GHz}) - \Delta ) $ is the
percent fraction of 31~GHz emission not due to free-free emission.

$^d$ The 5~GHz flux for NGC~1360 is contaminated by a point source. 

$^e$ Upper limits are 3~$\sigma$. 
\end{flushleft}

\end{table*}

\subsection{CBI polarization upper limits.}

Four PNe were observed by the CBI in polarization during November 2002
and June 2005: NGC7009, NGC1360, NGC246 and NGC7293 (the Helix
nebula). The polarization capability of the CBI has been demonstrated
with deep observations of CMB \citep{rea04} and also in H{\sc ii}
regions \citep{car05,dic06}. Details of the polarization calibration
and data reduction procedures can be found in the aforementioned
papers.

Natural-weighted maps of Stokes $Q$ and $U$ were made using the {\sc
difmap} package and combined to make polarization intensity maps,
$P=\sqrt{Q^2+U^2}$.  All maps were consistent with noise -- no
significant polarization was detected in any of the four
PNe. Table~\ref{tab:pol} summarizes the results and includes the
$3\sigma$ upper limit on the polarization fraction, $p=P/I$, derived
from the data.

The 99\% confidence level (CL) upper limit on the polarization
fraction was calculated assuming the $Q$ and $U$ maps have normal
distribution with zero mean and a standard deviation of $\sigma$. The
polarized intensity, $ P = \sqrt{Q^2 + U^2}$, follows the Rayleigh
distribution $f(P)$. The CL $F(P_{u})$ associated to an upper limit
$P_{u}$ on the polarized intensity is thus $ F(P_{u}) = \int_0^{P_{u}}
f(P) dP $, and a CL of 99\% corresponds to $P_{u} \approx
3\sigma$. Finally the upper limit on the fraction of polarized
intensity is $P_{u} / I_\mathrm{max} $, where $I_\mathrm{max}$ is the
peak total intensity (which in units of Jy~beam$^{-1}$ is equal to the
source flux density for unresolved sources).

%
\begin{table}
\centering
\caption{CBI 31~GHz polarisation measurements for four PNe. Upper
  limits are $99\%$ C.L..} 
\label{tab:pol}
\begin{tabular}{lr}
\hline
PNe             & Polarisation fraction \\      
\hline 
NGC~7009         &$<8.5~\%$ \\
NGC~1360         &$<8.5~\%$ \\
NGC~246          &$<17~\%$ \\
NGC~7293         &$<2.4~\%$ \\
\hline
 \end{tabular}
\end{table}
\normalsize

Strong radio polarisation is not expected for PNe since the emission
is typically dominated by free-free emission, which is intrinsically
unpolarised. However, the excess anomalous component could perhaps be
significantly polarised, depending on the physical mechanism producing
the emission. The most constraining and interesting result here is for
the Helix nebula where significant excess emission was observed at
31~GHz with the CBI \citep{cas04}. Since free-free radiation is not
polarised, we can limit the polarisation fraction on the excess
emission in the Helix to $<$3.8--12~\%, for 36--80~\% free-free emission
at 31~GHz.

\section{SIMBA  observations} \label{sec:simba}

SIMBA, at SEST\footnote{the SEST is situated on La Silla in Chile}, is
a 37-channel bolometer array, operating at 1.2~mm (250~GHz). The
half-power beam-width of a single element is 24$''$.  We observed each
PN with a scanning speed of 80~$''$~s$^{-1}$, in 51 steps of 8~$''$
orthogonal to the scan direction, thus obtaining $900 '' \times 408
''$ maps. The SIMBA scans are reduced with the standard recipe for
point sources using the MOPSI package written by Robert Zylka (IRAM,
Grenoble).  Flux calibration is carried out by comparison with Uranus
maps. We extracted the calibration flux densities of Uranus by
integrating the Uranus specific intensity map in a circular aperture
1~arcmin in radius about the centroid of an elliptical Gaussian fit. A
residual sky background was estimated from the median intensity in a
ring 1~arcmin wide and immediately surrounding the flux extraction
aperture.

We observed 35 PNe during 3 different observing runs in 2001 and
2002. Flux densities were measured by integrating each PN specific
intensity map in a circular aperture 1~arcmin in radius about the
centroid of an elliptical Gaussian fit to the PN image, and
subtracting a residual sky background, as in the case of Uranus.

The uncertainty assigned to each individual integration is the rms
noise in a circular annulus surrounding our extraction aperture (and
taking into account correlated pixels). Up to three integrations from
different days on individual objects were combined to reduce the
uncertainties.

A $\sim$10\% calibration uncertainty affects SIMBA flux densities. The
systematic error comes from the uncertainty in the brightness
temperature estimate of Uranus and the uncertainty in the
determination of telescope parameters, such as the aperture
efficiency. A fractional systematic uncertainty with a standard
deviation of 10\% was therefore added in quadrature to the combined
statistical uncertainties derived from the aperture photometry.

%

\begin{figure}
\begin{center}
\includegraphics[width=0.8\columnwidth,height=!]{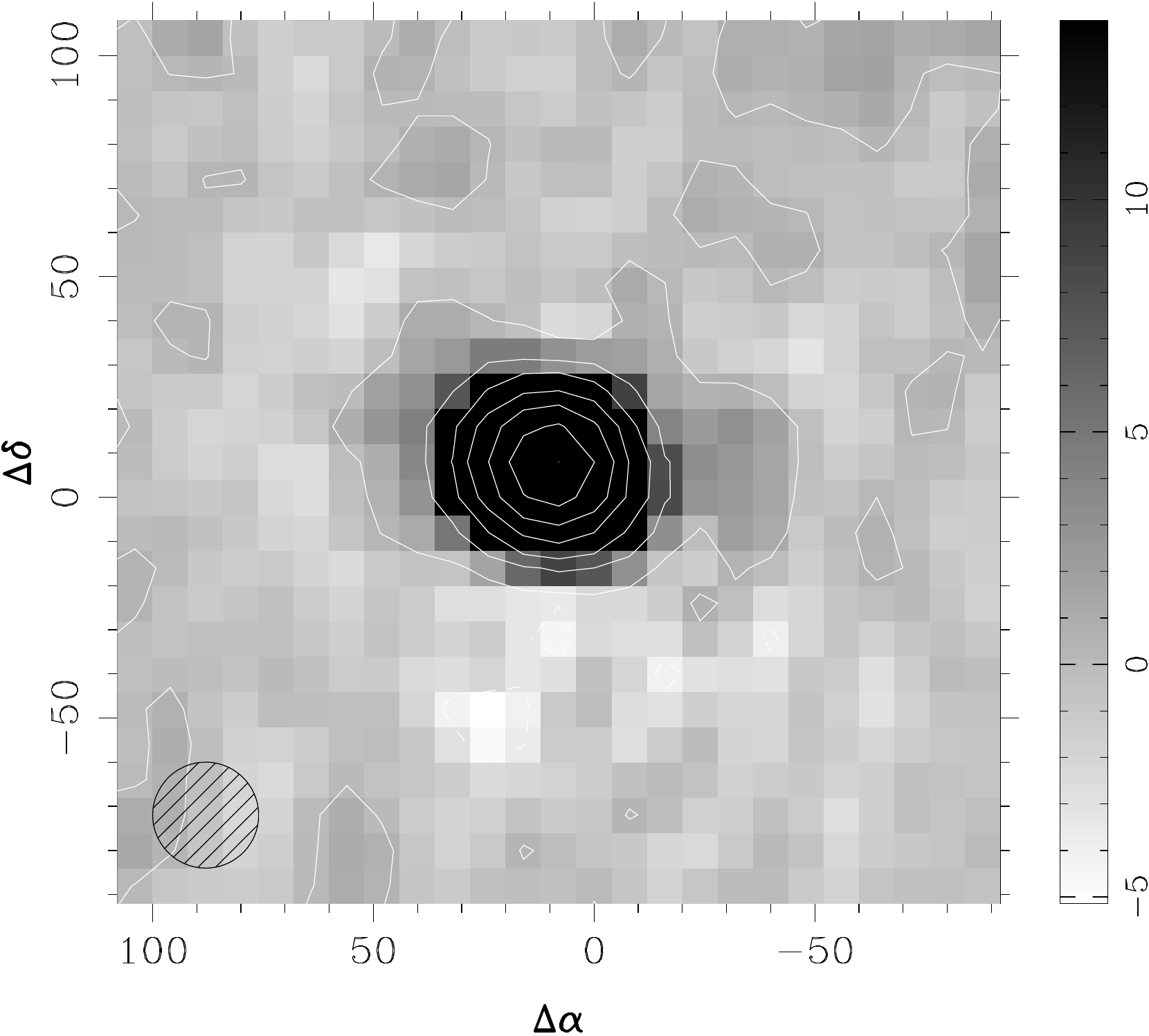}
\end{center}
\caption{\label{fig:simbaexample_6302} SIMBA map of NGC~6302. The gray
  scale is in units of MJy/sr, and the peak intensity is 138~MJy/sr
  (or 2.1~Jy/beam, for SEST's circular beam with a FWHM of 24~arcsec
  at 250~GHz). $x$-axis and $y$-axis show offset J2000 RA and DEC in
  arcsec. Contour levels at -4.9, -3.6, 0.2, 7.5, 19.5, 37.4, 62.4,
  95.7, 138.4, MJy~sr$^{-1}$. The SIMBA beam is indicated on the
  lower-left corner of the map.}
\end{figure}

Fig.~\ref{fig:simbaexample_6302} illustrates our SIMBA images.  All
the SIMBA images can be approximated to elliptical Gaussians. The only
case where the nebular solid angle $\Omega_N$, as inferred from the
Gaussian fits, is larger than 1.5 times the SIMBA beam, $\Omega_S$, is
NGC~6369, with $\Omega_N = 1.6 \Omega_S$. In the case of NGC~6302
(shown on Fig.~\ref{fig:simbaexample_6302}) low level emission can be
resolved by SIMBA.  The NE-SW low-level extension in
Fig.~\ref{fig:simbaexample_6302} follows the bipolar axis of
NGC~6302. The stripes and negatives in the SIMBA maps are proportional
to the object flux densities. They are artifact inherent to the
technique of fast scanning, which involves a high-pass filter to
cancel sky emission. These artifacts are also present on the Uranus
calibration images, and are taken into account by the flux calibration
procedure.

%
%
%


\section{CBI-SIMBA comparison} \label{sec:comp}

As summarised in Table~\ref{table:master} we have obtained 31~GHz and
250~GHz flux densities for a total of 21 object, of which 3 have
$\chi^2_\mathrm{CBI} >2$. Optically thin free-free emission at 31~GHz
can be extrapolated to 250~GHz with a spectral index of $\alpha =
-0.15$. We find that 20 out of 21 objects have 250~GHz data
consistent, within 1~$\sigma$, with a deficit over the free-free level
extrapolated from 31~GHz. Only 4 objects have $<1$~$\sigma$ excesses
at 250~GHz rather than a deficit (i.e. $\Delta > 0$, in the notation
of Table~\ref{table:master}). The 250~GHz deficit is significant in 9
PNe, at 3~$\sigma$. The only significant detection of an excess at
250~GHz is in M~2-9, with a 5~$\sigma$ excess at 250~GHz.

%

The CBI-SIMBA spectral index, averaged over the 17 measurements of
$\alpha_{31}^{250}$ listed in Table~\ref{table:master} that have
$\chi^2_\mathrm{CBI} < 2$, and excluding M~2-9, is $\langle
\alpha_{31}^{250}\rangle = -0.43 \pm 0.03$, with a weighted rms
scatter of 0.14. Thus, the observed $\langle \alpha_{31}^{250}
\rangle$ is significantly different from the free-free value.

\begin{figure*}
\begin{center}
\includegraphics[width=0.9\textwidth,height=!]{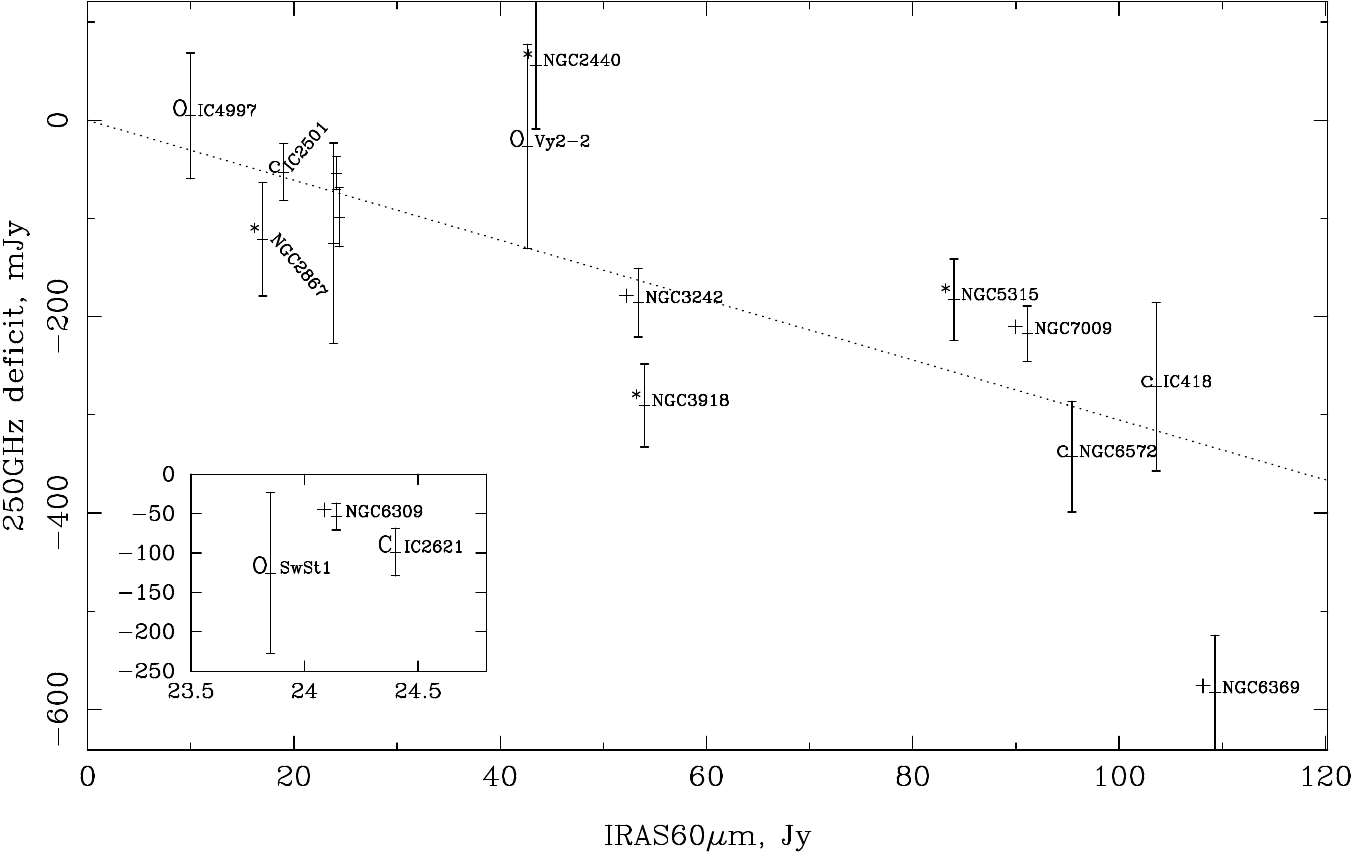}
\end{center}
\caption{\label{fig:excess} Deficit at 250~GHz from the expected level
  of free-free emission extrapolated from the CBI measurements,
  excluding M~2-9 for clarity. The $y-$axis shows the difference
  between the SIMBA flux density and the free-free level extrapolated
  from 31~GHz, and the $x$-axis shows {\em IRAS}~60$\mu$m flux
  density. The dotted line is a fit to the data.}
\end{figure*}

We searched for a correlation between the 31~GHz excess and the mid-IR
continuum due to VSGs, as observed in the anomalous CMB
foreground. Unfortunately the {\em IRAS}~12$\mu$m band is contaminated
by strong ionic lines as well as free-free emission, while
ground-based 10$\mu$m spectra available from the literature are
extracted from varying apertures. We can still compare the 31~GHz
excess to the {\em IRAS}~25,~60, and 100~$\mu$m flux densities. The
{\em IRAS}~60$\mu$m band turned out to give the best correlation with
the 250~GHz deficit. Fig.~\ref{fig:excess} plots the difference
between the expected level of free-free emission at 250~GHz,
extrapolated from 31~GHz, against {\em IRAS}~60$\mu$m. The 10$\mu$m
dust emission features compiled from the literature are indicated by
'C' for PAHs, 'c' for SiC, 'O' for silicates, '+' for weak and
featureless continuum, and '*' when no data are available \citep[as
  in][]{cas01}. A straight line fit to the data, required to cross the
origin, gives a dimensionless slope of $(-3.05\pm0.18)~10^{-3}$. The
cross-correlation with {\em IRAS}~60$\mu$m could simply reflect an
overall scaling of the flux densities with nebular luminosity. The
cm-excess does not seem related to the 10~$\mu$m dust emission
features.


We also calculated the 250~GHz deficit when extrapolating the
free-free level from the 5~GHz data. The resulting deficits are still
significant, albeit for a reduced number of objects. The reduced
significance may be due in part to the 10\% calibration uncertainty of
the 5~GHz data, but also to a 5-31~GHz spectral index greater than the
optically thin value (see Sec.~\ref{sec:cbi}). In this case a straight
line fit to the 250~GHz deficit as a function of {\em IRAS}~60$\mu$m
band flux density, for the same objects as for the 31~GHz
cross-correlation of Fig.~\ref{fig:excess}, gives a dimensionless
slope of $(-2.65 \pm 0.27)~10^{-3}$.

%
%

\begin{figure*}
\begin{center}
\includegraphics[width=\textwidth,height=!]{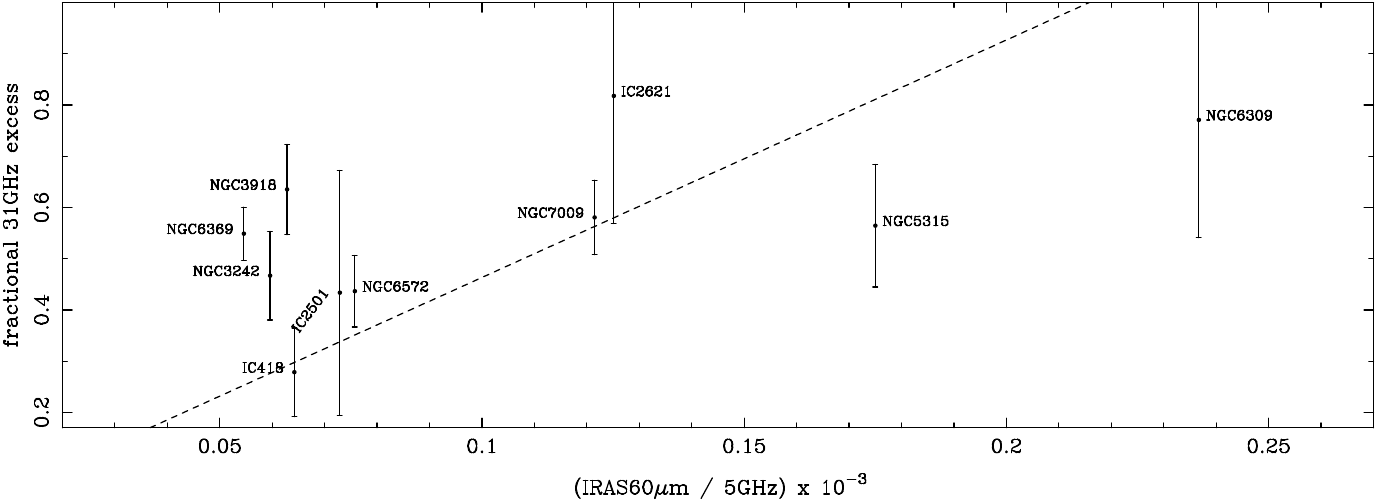}
\end{center}
\caption{\label{fig:fracexcess} The ratio of the 31~GHz excess
  (extrapolated from the 250~GHz measurement, assuming it is entirely
  due to free-free emission) over the total 31~GHz flux density,
  against the ratio of the IRAS~60~$\mu$m flux density over the 5~GHz
  flux density. The dashed line is a fit to the data that crosses the
  origin, with a slope of $4.54\pm1.5$ and reduced $\chi^2$ of 0.5. }
\end{figure*}

Relative to the total 31~GHz flux density, the excess emission over
the level of free-free emission extrapolated from 250~GHz represents
$\sim$0.3--0.7. In order to avoid induced correlations from scaling
effects, we plot in Fig.~\ref{fig:fracexcess} the ratio of the 31~GHz
excess to the total 31~GHz flux density against the ratio of the
IRAS~60~$\mu$m flux density over the 5~GHz flux density. We have
excluded from Fig.~\ref{fig:fracexcess} all objects with large
$1~\sigma$ uncertainties on the fractional excess ($>0.3$). The
remaining 10 objects, not including M\,2-9, have an average fractional
excess of $0.51\pm0.03$ with an rms scatter of 0.11.  Interestingly,
the fractional excess seems to correlate with the ratio of far-IR and
radio flux, as expected for a dust-related emission mechanism at
31~GHz.

We caution that the IR-radio slopes derived in this Section cannot be
directly compared with the slopes observed in the ISM at large. PNe
have much hotter dust temperatures \citep[~100--200~K, e.g.,
][]{cas01} than dark or cirrus clouds (~15--25~K).

%
%

\section{Comparison with previous work} \label{sec:previous}

\subsection{NGC~7027}

An important PN which is absent from our sample is northern NGC~7027,
the brightest at radio wavelengths. \citet{ter74} and \citet{hoa92}
constructed the SED of NGC~7027, finding its radio spectrum follows
free-free emission. In this respect NGC~7027 is similar to NGC~6302
(discussed in Sec.~\ref{Sec:NGC6302}), another high-excitation PN with
a massive progenitor. Both are surrounded by molecular envelopes with
copious amounts of dust. Although any cm-excess appears to be
negligible in NGC~7027 and NGC~6302, the mm-wave data could be
affected by either molecular lines or cold dust, and definitive
conclusions in these objects require further analysis. NGC~7027 and
NGC~6302 are reminders that PNe form a heterogenous class.

\subsection{Previous mm-wave flux densities from the literature.}

Deviations from standard free-free spectra in the radio continua of
PNe were already present in the SEDs obtained by \citet{hoa92},
although not considered significant. It can be seen in Fig.~1 from
\citet{hoa92} that NGC~6572, NGC~6302, NGC~6537 and NGC~6543 show
deficits at $200-300~$GHz. We confirm that our 1.2~mm flux densities
are consistent with those measured at 1.1~mm by \citet{hoa92}, within
2~$\sigma$, for the four objects we have in common: NGC~6572,
NGC~6302, M2-9, and NGC~6537. These objects are studied in detail
below.

\subsection{Preliminary mm-wave flux densities from SEST.}

The 250~GHz deficit over the expected level of optically thin
free-free emission was originally found by one of us (L.-\AA.N) with
heterodyne and single pixel bolometer observations at SEST. But in
pointed observations the measurement of flux densities is hampered by
pointing uncertainties. Until the development of bolometer arrays such
as SIMBA, designed to produce images of the sky, it was difficult to
ascertain whether the 250~GHz deficits were not due to pointing errors
\citep[which may also have been the problem in][]{hoa92}.

The SEST heterodyne observations observations were performed between
1995 and 1996 with the 15 m SEST. The data were obtained using
simultaneously two single-channel SIS receivers at 99 and 147 GHz with
typical system temperatures of 150 K (SSB) above the atmosphere.  The
telescope beamwidth is 50$\arcsec$ at 99 GHz and 34$\arcsec$ at 147
GHz (FWHM). Two acousto-optical spectrometers (AOS) were used
simultaneously with bandwidths of about 1 GHz, channel separations of
about 0.7 MHz, and resolutions of about 1.4 MHz. The telescope was
used in a dual beam switch mode with the source alternately placed in
each of the two beams, a method that yields very flat baselines, and
allows the determination of continuum levels as the offset of the
baseline from the zero level . The beam separation was about 11$
\farcm$5.  The flux densities were determined from the chopper-wheel
corrected antenna temperature, $T_{\rm A^*}$, using aperture
efficiencies, ($\eta_{\rm A}$), determined from observations of
planets, of 0.60 and 0.52 at 99 and 147 GHz, respectively, and are
estimated to be accurate to within 10 \% (one sigma). The pointing was
checked during the observations using SiO maser sources, and the
pointing accuracy is estimated to be better than $\pm 3\arcsec$. 


The SEST single pixel bolometer observations were performed between
1993 and 1995, using a single channel bolometer developed at MPIfR,
operating at a wavelength of 1200~$\mu$m (i.e., as for
SIMBA). Beam-switching was done with a focal-plane chopper having a
horizontal beam throw of 70$''$. The observations were done in the
ON-OFF mode, alternating the source between the two beams after 10 s
of integration time. Pointing was checked on nearby quasars and was
found to be better that 3$ \arcsec$ rms. The atmospheric opacity was
determined frequently through skydips. Uranus was used as a calibrator
source, and the estimated uncertainty in intensities are about 10\%
(one sigma). For single-pixel observations we report point-source flux
densities, with uncertainties derived from the rms dispersion of all
ON-OFF pairs.  The maps were obtained by scanning the telescope in
azimuth at a rate of 8 $\arcsec$ per second, with adjacent scans
separated by 8$\arcsec$ in elevation. The maps were reduced using the
NOD2 software \citep{has74} adapted for SEST. Flux densities were
extracted with Gaussian fits.

We summarise in Table~\ref{table:prevflux} the measurements from the
preliminary SEST observations, both from the single pixel bolometer
and the heterodyne backends. In general the previous pointed
observations at 230~GHz are within 3~$\sigma$ with the 250~GHz values
from Table~\ref{table:master}. The only exceptions are Hb~5, NGC~6537
and NGC~3132, where the pointed data gives lower flux densities
(consistent with a pointing error, or with nebular extensions larger
than the beam), and NGC~7009, where the difference with the 230~GHz
map is at 3.1~$\sigma$. The concordance of the whole dataset on
NGC~6369 is particularly comforting, and will be commented further
below.

\begin{table}
\centering
\caption{Previous flux densities from SEST. Values are reported in
  mJy. Systematic uncertainties are not included, and are assumed to
  be of $\sim30\%$.}
\label{table:prevflux}
\begin{tabular}{lrrrr}
\hline
PN              & \multicolumn{2}{c}{Single pixel}    & \multicolumn{2}{c}{Heterodyne}                  \\      
                &   ON-OFF               &   map      &  147~GHz       &  99~GHz                       \\       \hline 						       
NGC3132     	&	$ 30.1	\pm15.0 $&            &          &   \\ 
NGC3242     	&	$180.2	\pm 9.0 $&            &          &   \\ 
NGC3918     	&	$149.4	\pm 7.9	$&	      &	497.7 &	342.6\\ 
NGC5315     	&	$136.4	\pm 9.9	$&	      &	      &	     \\ 
NGC6072     	&	$16.6	\pm 9.4	$&	      &	      &	     \\ 
M2-9        	&	$276.8	\pm 11.2$&	      &	107.4 &	149.7\\ 
NGC6302     	&	$2091.9	\pm 13.7$&	1890  &	1890  &	2054 \\ 
NGC6369     	&	$396.5	\pm 12.9$&	485   &	783   &	1094.\\ 
M1-26       	&	$185.7	\pm 15.6$&	      &	368.1 &	487.7\\ 
Hb5         	&	$89.7	\pm 8.6	$&	      &	250.2 &	337.4\\ 
NGC6537     	&	$80.1	\pm 11.1$&	      &	      &	     \\ 
NGC6572     	&	$670.8	\pm 12.1$&	434   &	843   &	882.1\\ 
NGC6790     	&	$154.8	\pm 9.4	$&	      &	120.6 &	176.0\\ 
Vy2-2       	&	$94.2	\pm 15.2$&	      &	      &      \\ 
NGC7009     	&	$178.2	\pm 16.5$&	279   &	231   &	379.6\\
\hline
 \end{tabular}
\end{table}
\normalsize

\subsection{Absence of radio PN haloes.}

Neither the CBI nor the SIMBA datasets show evidence for radio
haloes. When compared with the point-source models, none of the
core-halo models that we tried to fit the CBI visibilities gave
improved $\chi^2_\mathrm{CBI}$ values. Extended haloes, about 4~arcmin
in diameter and larger, have indeed been observed in a few PNe
\citep{cor03}. But the optical H$\alpha$ + [N\,{\sc ii}] haloes are
extremely faint. The halo intensities are of order 1/1000 the values
in the central regions that dominate our radio measurements. The
integrated halo line fluxes are about 1/100 the fluxes of the compact
regions.

\subsection{Is the PN cm-wave excess seen in H\,{\sc ii} regions?}

As natural comparison objects to PNe, H\,{\sc ii} regions are also
photoionised nebulae with thermal continua. Does their radio continua
also bear the cm-excess? H\,{\sc ii} regions are extended and
complex objects; the measurement of precise radio spectral energy
distributions is difficult. Single-dish maps at frequencies higher
than $\sim$14~GHz require cancelling the sky emission with either fast
scanning (as for SIMBA) or differencing, while interferometric data
are affected by flux loss and a wavelength-dependent
$(u,v)$-coverage. Another difficulty with mm-wave continuum data of
H\,{\sc ii} regions is the possibility of molecular line emission
from surrounding or coincident molecular material \citep{sut84}.

Notwithstanding these difficulties, \cite{dic07} reported that the
radio SEDs of six H\,{\sc ii} regions constructed from CBI
observations and literature data, are compatible with free-free
emission. They constrain the possibility of spinning dust, and note a
3~$\sigma$ 31~GHz excess in G284.3-0.3 similar to spinning dust in its
narrow profile.

But the SEDs of H\,{\sc ii} regions at mm-wavelengths are not
established. A close look at the Boomerang and SEST data reported by
\citet{cob03} reveals that H\,{\sc ii} regions may be affected by the
same 250~GHz deficit seen in PNe. The SED of RCW38 can be constructed
with the Boomerang data \citep{cob03} and the CBI data
point\footnote{We assigned 10\% uncertainties to the CBI data point to
  allow for differences in the photometric extraction procedures: the
  CBI flux densities are obtained by Gaussian fitting, while
  \citet{cob03} used a circular photometric aperture encompassing most
  of the emission}.  We fit the 31--400~GHz SED of RCW38 with two
power laws. The best fit index at high frequencies is $\alpha =
+3.58\pm0.69$, which is consistent with grey body emission (with an
emissivity index of $\beta = 1.58\pm0.69$). The low frequency data are
fit by $\alpha = -0.22^{-0.08}_{-0.38}\pm0.14$ (where we have
indicated the $\Delta \chi^2 = 1$ limits). Given the uncertainties
such an index over 31--90~GHz is consistent with the indices seen in
PNe.


%

%

The H\,{\sc ii} region SEDs published by \citet{mal79} also cover the
mm-wave regime. The 10--100~GHz index is $\alpha \sim -0.25$ in W51,
and is $\alpha \sim -0.36$ in W3. Uncertainties are difficult to
estimate from the figures in \citet{mal79}, but our point here is that
given the existing data, H\,{\sc ii} regions may well show the same
cm-excess as PNe.

%

\section{Candidate emission mechanisms} \label{sec:candidates}

In this Section we speculate on the emission mechanisms that could
explain the observed PN SEDs. We take the case of NGC~6369 as an
example, since it is the object with the most significant 250~GHz
deficit. We assume a distance for NGC~6369 of 330~pc, and a nebular
diameter of 38~arcsec \citep[from optical images, as compiled
  by][]{ack92}. Refer to Fig.~\ref{fig:NGC6369}c and
Sec.~\ref{sec:needles} for a comparison of the data and a pure
free-free spectrum. The free-free continua were calculated with the
exact Gaunt factors for hydrogen.

\subsection{A synchrotron component?} \label{sec:sync}

Assuming the 250~GHz continuum is due only to free-free emission and
possibly to the thermal dust Rayleigh-Jeans tail, any synchrotron
component should be very steep above 31~GHz, with a spectral index
steeper than $\alpha_{31}^{250}\rangle$, or about $-0.4$
(Table~\ref{table:indices}). This is in contrast with $\langle
\alpha^{31}_5 \rangle$, which is essentially flat.


Synchrotron self absorption is not expected in PNe. But the flat index
at low frequencies could be explained by free-free absorption with a
turn-over frequency at ~20--30~GHz. For a typical emission measure of
$10^6$~cm$^6$pc$^{-1}$, a turnover frequency of 40~GHz requires a
free-free screen with an electron temperature $T_e < 600~$K for the
bulk of the nebula.


\subsubsection{Background}

Synchrotron radiation was sought in PNe during the 1960s as a means to
detect the magnetic field required by \citet{gur69} in the theory of
magnetic shaping \citep[see also, e.g.,][]{pas85,che94}. The evidence
put forward by \citet{gur69} was based on an UV excess in photospheric
spectra, and by the mention of non-thermal emission being observed in
a few objects. However, only the case of NGC~3242 appears to be backed
by a reference. Although \citet{men65} report a non-thermal spectrum
for NGC~3242, their data is not confirmed by the flux densities given
in \citet{cal82}, \citet{con98} and \citet{mil82}.

The possibility of non-thermal emission from the interaction of a
magnetised and fast wind with dense knots has been investigated by
\citet{jon94,jon96}, in diverse astrophysical contexts. They expect
synchrotron emission to arise at the shocked interfaces between knots
and fast wind. Synchrotron should be strongest along the stretched
magnetic field lines around the perimeter of the clumps, and increase
at the time of knot disruption.

An interesting alternative has been proposed by \citet{dga98}. In the
specific case of PNe, \citet{dga98} argue that simultaneous cosmic ray
acceleration and magnetic field enhancement could occur in the
turbulent mixing layers at the sides of the knots and their turbulent
wakes.  They find that such synchrotron emission could be detectable,
at a level of 1~mJy at $\nu \sim 1$GHz, in the case of hydrogen
deficient nebulae, where the level of free-free emission is reduced.
\citet{dga98} claim the detection in PN A~30 of faint (0.1~mJy),
compact (3~arcsec), non-thermal and circumstellar emission at
8~GHz. But the data have not been published.

There is thus no data available to back the synchrotron hypothesis for
the cm-excess. Recently \citet{coh06} have reported the
detection of two non-thermal hot spots in the faint PN surrounding the
OH/IR star V1018~Sco (with spectral indices of $-0.85\pm0.01$ and
$-0.95\pm0.11$ over 5 and 10~GHz). \citet{coh06} attribute the
ionisation of this object to the interaction between a fast AGB wind
that has been recently turned on and is overrunning a precursor slow
wind. It is very different from the photoionised nebulae which make
the focus of this work.

Magnetic fields have however been detected by \citet{mir01} in the
young PN K~3-35, through the detection of a circularly polarised OH
maser line at 1.665~GHz, implying field intensities of order $\sim
1$~mG.  Fields of $\sim$1~mG are in the range of the theoretical
predictions of \citet{dga98}, based on the equipartition of the
magnetic field and fast wind energies.

\citet{vle06} found water masers in the proto-PN W43A, whose
polarization traces the precession of a strong collimating magnetic
fields, $>1$~mG on 1000~AU scales. \citet{bain03,bain04} provide
further detections of magnetic fields in proto-PNe.  \citet{sab07}
observed the nebular sub-mm dust polarization concomitant to a
pervasive $\sim1$~mG field.

In fact, the energetic requirements of synchrotron radiation are
easily met in PNe. Given a magnetic field and a synchrotron index
$\alpha$ observed over a frequency range from $\nu_1$ to $\nu_2$,
the total energy $U_e$ in cosmic rays can be estimated from
\citep{mof75}:
\begin{eqnarray}
U_e &=& N_T  \frac{E_2^{2-p} - E_1^{2-p}}{2-p}, ~\mathrm{where},  \\
N_T &=& L_\mathrm{obs} / \left[ C_3  B^2 \frac{E_2^{3-p} -
    E_1^{3-p}}{3-p}  \right], ~\mathrm{and} \\
E_i^2 &=&   \frac{\nu_i 4 \pi m_e^3 c^5}{3 e B} , ~\mathrm{i=1,2},  \\
p & = &  1 -  2 \alpha, ~\mathrm{and}~ C_3 = 2.4~10^{-3}~\mathrm{cgs}.
\end{eqnarray} 
The observed synchrotron luminosity in NGC~6369, between 0.1~MHz and
31~GHz, is $L_\mathrm{obs} = 8.7~10^{30}$~erg~s$^{-1}$, as obtained by
integrating a power law spectrum with $F_\nu(31~\mathrm{GHz}) =
1.5~$Jy and $\alpha = -0.3$. We find that, for a 1~mG field
permeating the whole nebula, the total energy in cosmic rays required
to account for $L_\mathrm{obs}$ is $\sim~5~10^{42}$~erg. Since the
lifetime of a typical nebula is $10^4$~yr, such an amount of
cosmic-ray energy requires a stellar luminosity of $\sim
4~10^{-3}$~L$_\odot$, much less than the observed tip-of-the-AGB
luminosity of $10^4$~L$_\odot$\footnote{Synchrotron radiation from the
  nebular knots taps energy from the fast stellar wind, which in these
  hot stars is accelerated by radiation pressure coupled to the gas
  through ionic lines \citep[e.g.][]{pau88}. The radiation - fast wind
  coupling must be very effective since observations give
  $P_\mathrm{rad} \sim P_\mathrm{wind}$.}

It is interesting to compare the above energetic requirement and the
calculation in \citet{dga98}. In their idea synchrotron should stem
from the whole ensemble of globules and their overlapping turbulent
wakes, encompassing a linear size $R_\mathrm{ens}$.  \cite{dga98}
assume equipartition between magnetic field, cosmic ray electrons, and
fast wind.  Their Eq.~9 gives a synchrotron luminosity of only
$4.9~10^{28} (R_\mathrm{ens}/5~10^{15}\mathrm{cm})^3$~erg~s$^{-1}$.
Accordingly \citet{dga98} predict synchrotron flux densities 1/100 of
the values required for the cm-excess. But a value of
$R_\mathrm{ens}=5~10^{15}$~cm is actually very small for a typical
PN. As seen in the Helix nebula, the ensemble of knots covers at least
1/3 of the nebular solid angle, and the nebulae considered in this
work are typically $\sim$0.1~pc. Given that the \citet{dga98}
synchrotron luminosity scales as the cube of $R_\mathrm{ens}$, a mere
increase to $2.5~10^{16}$~cm, still much less than 0.1~pc,
approximates the cm-excess.

\subsubsection{NGC~6369}

\begin{figure}
\begin{center}
\includegraphics[width=\columnwidth,height=!]{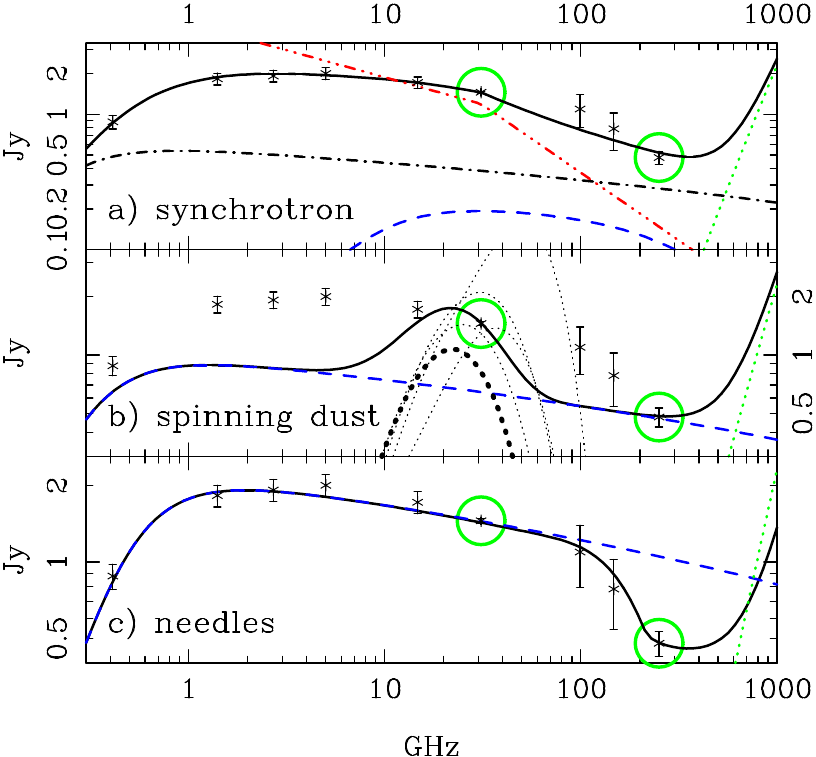}
\end{center}
\caption{\label{fig:NGC6369} SED of NGC~6369. We plot flux density in
  Jy ($y-$axis) as a function of frequency in GHz ($x-$axis). The CBI
  and SIMBA datapoints are indicated by circles.  {\bf a- Synchrotron}
  The solid line includes a synchrotron component with a break at
  31~GHz (dash double dotted), a core free-free component at $10^4$~K
  (dash dotted), a sub-mm modified black-body (dotted), and the
  emission expected from the free-free screen, chosen here to be at
  20~K (dashed). {\bf b- Spinning dust} The dashed line is a free-free
  component at 7400~K, with $N_e = 2292$~cm$^{-3}$. A spinning dust
  component for the ``WNM'' case of \citet{dl98b}, is shown in thick
  dotted line and contributes to the total flux density shown in solid
  line. The other ISM environments considered by \citet{dl98b} are
  shown in dotted line, scaled by the same factor as the `WNM'
  case. The low frequency data cannot be reconciled with the spinning
  dust models. {\bf c- Needles} The solid line includes extinction
  from very cold metallic needles, the dashed line is the free-free
  component, and the dotted line is the sub-millimetre tail of a
  modified black body, attenuated by gray needle extinction. Since
  needle absorption does not extend appreciably below 100~GHz, the
  dashed line indicates nebular emission had it been pure free-free. }
\end{figure}

Given the detection of magnetic fields in PNe, we did the exercise of
modelling the SED of NGC~6369 including a synchrotron component
(Fig.~\ref{fig:NGC6369}a). The core free-free emission, which
represents the bulk of the 250~GHz flux density, is approximated as a
uniform slab at $10^{4}$K. The synchrotron component is modulated by a
cold screen, with a power-law distribution of opacities (as in the
case of M~2-9, see Sec.~\ref{sec:M2-9}), which is required for
NGC~6369 to fit the curvature of the $<30$~GHz data\footnote{This cold
  free-free component cannot, of its own, explain the cm-excess,
  because the lowest possible free-free index is $\alpha = -0.175$ at
  1000~K (see Fig.~\ref{fig:ffindex}), and because the nebulae, as
  cold blackbody radiators, would have to be much larger to reach the
  observed $<5$~GHz levels}. The cold screen opacity includes both
free-free and bound-free contributions, although the latter is
negligible at frequencies where the screen opacity is $\sim$1.

%

We could only obtain a fit to the data by including a synchrotron
break at 31~GHz. For a single power law to account for the whole SED,
it would need to join the CBI and SIMBA data points, so that free-free
emission would be altogether negligible compared to synchrotron. Thus,
the best fit we obtained with a single power law has an index of
$-0.57$, and a screen at 15K, with a power law distribution of
opacities (see Sec.~\ref{sec:M2-9}). This model fits only marginally
the low frequency data. Note that the need for a high frequency break
is alleviated by the inclusion of a spinning dust component (see
Sec.~\ref{sec:spin}).


Any free-free screen with a temperature below $\sim$~1000~K can fit
the data, but the electron density in the screen is an increasing
function of temperature. At a fixed screen temperature of
$T_\mathrm{screen} = 500$~K, $\langle N_e(\mathrm{screen}) \rangle =
1221\pm291$~cm$^{-3}$ and $N_e(\mathrm{core}) = 1446\pm507$~cm$^{-3}$,
while at $T_\mathrm{screen} = 10$~K all of the 250~GHz flux density is
due to the core free-free emission, $N_e(\mathrm{core}) =
2404$~cm$^{-3}$ and $\langle N_e(\mathrm{screen}) \rangle =
33\pm20$~cm$^{-3}$. Perhaps an advantage of the lower temperatures is
that the low-frequency synchrotron index can be somewhat steeper; at
1000~K the index is fixed at $\sim -0.1$ by the 5-31~GHz data.

For integrated flux densities the 2-D opacity fields can be modelled
following \citet{cal91}.  Section~\ref{sec:M2-9} on M~2-9 gives more
detail on the procedure, as well as an eloquent example of an opacity
profile. Fig.~\ref{fig:NGC6369}a shows the fit to the SED of NGC~6369
with a fixed low-frequency synchrotron index of $\alpha_\mathrm{LF} =
-0.4$, a fixed high-frequency synchrotron index of $\alpha_\mathrm{HF}
= -1.0$, a screen at 20~K, wich has an average density of
$574\pm88$~cm$^{-3}$ and a power-law opacity index $\gamma = 6.0$ (see
Sec.~\ref{sec:M2-9}). A modified blackbody with emissivity index
$\beta = +2$, required to pass through the 100- and 60-$\mu$m {\em
  IRAS} flux densities, corresponds to $T_\mathrm{dust} = 52~$K and is
negligible at $<250~$GHz.

%


\subsubsection{Predictions of the synchrotron interpretation}

The main difficulty for the synchrotron interpretation is thus the
need for a cold free-free screen. Screen temperatures of 1000~K are
reminiscent of the ORL temperatures, but in the bi-abundance model the
ORL gas only accounts for a very small fraction of the nebular plasma,
while the 1000~K synchrotron screen represents about half the total
nebular mass. In contrast, the average electron density of the very
cold screen is a factor of $\sim 1/100 - 1/1000$ less than the bulk
nebular density. Such a 10--20~K screen could perhaps be found in the
residual ionisation of dense and largely neutral globules exposed to
the central star wind, where \citet{dga98} expect cosmic ray
acceleration.

The cold free-free opacity profile, as a function of a 1-D nebular
solid angle parameter $\omega$ (Sec.~\ref{sec:M2-9}), must be a power
law to attenuate the synchrotron spectrum and mimick a free-free
spectrum in emission below $\sim$31~GHz. Such opacity profiles occur
naturally in PNe with a large system of globules, which seems to be
the case in every object observed at sufficient resolution. A
prototypical example is the Helix nebula, where knots are known to be
dusty \citep{mea92} and molecular \citep{hug02}, with physical
conditions approaching the predictions of \citet{dys89}: temperatures
of $T \sim 10$~K, proton densities of up to $\sim 10^6$~cm$^{-3}$. If
the bulk of carbon in the knots is photoionised by the stellar UV
photons with energies $<16~$eV, then the electron density in globules
is about $10^2$~cm$^{-3}$. The C\,{\sc i} free-free continuum could
thus provide the cold free-free screen required to attenuate in-situ
the synchrotron background. If so, we except concomitant C\,{\sc i}
radio recombination lines at the sites of synchrotron emission.

Synchrotron emission should be clumpy as well. The cometary globules
providing the cold free-free screens must be coincident with the
synchrotron emitting regions.  The overlap between cold free-free and
synchrotron regions is natural in models where synchrotron emission
arises in the interaction between the fast central star wind and the
clumps \citep{jon96,dga98}. In the clumpy model the free-free
opacities should also include a filling factor ($f$, the fraction of
nebular volume occupied by the clumps). Thus the opacity for a nebula
with depth $L$ is $\tau = f N_e N_i \kappa_\mathrm{ff} L$, where the
free-free absorption coefficient is $\kappa_\mathrm{ff}$. Values of
$f\sim 1/10$ are allowed by the above synchrotron models with a factor
of 2-3 increase in the electron density $N_e$. Smaller values of $f
\sim 1/100$ can be obtained with shallower synchrotron backgrounds,
with low frequency indices of $\sim -0.2$.


%
%
%
%
%
%
%

If the cm-excess stems from the fast wind, then a relationship
exists between the cm-excess, the fast wind pressure $\dot{M}
v_\infty$, and X-ray emission from the shocked cavities predicted by
the interacting winds model \citep[e.g.,][]{kwo78}. But the sample of
PNe where diffuse X-rays have been detected is too small to allow a
cross-correlation with our sample. Of the PNe listed in
Table~\ref{table:master} only NGC~7009 and NGC~1360 are among the 13
PNe detected by {\em ROSAT}~PSPC \citep{gue00}. {\em XMM} and {\em
Chandra} have allowed to resolve the diffuse X-rays and isolate the
central star emission \citep{gue05}, adding only NGC~3242 to the
intersection with Table~\ref{table:master} (an up-to-date list of
X-ray PNe can be found at {\tt http://www.iaa.es/xpn/}).

The data on central star fast winds is also rather scarce, and no
trends are easily found. For example, NGC~6572, with a large cm-excess, has a high mass loss rate of $\dot{M} =
6~10^{-7}$~M$_\odot$~yr$^{-1}$ in a rather slow wind with $v_\infty =
1190$~km~s$^{-1}$ \citep[][]{mod93}. NGC~7009, also with a significant
excess, has a faster wind 2770~km~s$^{-1}$ but much smaller $\dot{M} =
2.8~10^{-9}$~M$_\odot$~yr$^{-1}$ \citep[][]{cer89}, which gives a ram
pressure $\sim$250 times weaker than in NGC~6572. IC~418, with a
rather small cm-excess, shares a similar fast wind as NGC~7009
\citep[$v_\infty = 940$~km~s$^{-1}$ and $\dot{M} =
  2.8~10^{-9}$~M$_\odot$~yr$^{-1}$,][]{cer89}.

A prediction of the synchrotron interpretation for the cm-wave
continua of PNe is that the nebular continuum should be polarized at
$\sim 60~$GHz. In order to calculate the expected level of synchrotron
polarization we assume the free-free screen is also embedded in a
pervasive nebular magnetic field of 1mG, and make use of the formulae
given in \citet[his Eq.~3.78]{pac70}. For a synchrotron spectral index
of $-0.5$, the fraction of intrinsic (unabsorbed) polarization is $\Pi
= 69\%$, in the case of a uniform magnetic field. Including a cold
nebular screen gives $\Pi = 69\%$ at 250~GHz, decreasing to $\Pi \sim
3-15\%$ at CBI frequencies. The total nebular emission includes an
unpolarized thermal component, which should account for most of the
250~GHz flux density to reconcile the continuum and recombination line
data in the literature. Thus, for a uniform magnetic field, the total
fraction of polarization $\Pi_T$ should reach a maximum of $\Pi_T \sim
20\%$ between 50 and 100~GHz.

But PNe are often very point-symmetric, so that the polarization
fraction on flux densities integrated across the whole nebulae should
vanish for perfect point-symmetry. Moreover, in the interpretation of
\citet{dga98}, the magnetic fields required for synchrotron emission
are amplified in turbulent wakes. This is a case of tangled magnetic
field, without net polarization. Thus, very high angular resolution
observations, at 50--100~GHz, are required to detect the polarization
predicted by the synchrotron interpretation.

In summary the synchrotron interpretation predicts that the radio
C\,{\sc i} recombination lines should be coincident, on angular scales
comparable to the cometary globules, with peaks in the continuum
polarization between 50 and 100~GHz.

\subsection{Is it spinning dust?} \label{sec:spin}


Spinning dust peaks at $\sim$30~GHz and is 30--70 times weaker at
5~GHz \citep[depending on environment,][]{dl98b}. If spinning dust
emission accounts for 30\% of the 31~GHz flux density, then
$\alpha^{31}_5 = 0.045$, which is consistent with $\langle
\alpha^{31}_5 \rangle$ within 2~$\sigma$ (see
Table~\ref{table:indices}). Spinning dust drops with a Boltzmann
cutoff above 30~GHz, and is $>300$ times weaker at 100~GHz. Could the
250~GHz deficit be explained by a 31~GHz excess due to spinning dust?
VSGs are found aplenty in PNe, and the proximity of a hot stars
warrants such VSGs will be charged by the photoelectric effect.

However, taking all the low-frequency measurements into account, we
find that the existing models for spinning dust cannot alone explain
the cm-excess in NGC~6369. Fig.~\ref{fig:NGC6369}b
illustrates that the low-frequency data have too flat a spectral index
to be explained by spinning dust. We used the spinning dust emissivity
of \citet[][{\tt http://www.astro.princeton.edu/\~draine/}]{dl98b}, for
the warm neutral medium case (WNM), which peaks at the lowest
frequency. The spinning dust emissivity, being electric-dipolar in
origin, stems from the Larmor power radiated by individual grains, and
has a very steep rise with frequency before peaking sharply at the
Boltzmann cutoff. Magnetic dust, proposed by \citet{dl99}, also with a
steep low-frequency rise, faces the same problem of not fitting the
low-frequency data.

One problem with spinning dust as a candidate emission mechanism is
that the current models give spectra that are too narrow in
frequency. Arbitrarily changing the frequency peak does not help,
since fitting the 5-10~GHz data would then miss 31~GHz. The spinning
dust emissivities of other ISM environments considered by
\citet{dl98b} all peak at slightly higher frequencies. Another problem
is that the predicted flux densities, calculated from the highest
diffuse ISM emissivities (the 'RN' case), barely reach peaks of 1~mJy
if scaled to nebular densities of $10^{4}~$cm$^{-3}$.  

The PN environment may nonetheless lead to higher spinning dust
emissivities, especially in the dense cometary knots, and the low
frequency levels could be explained by a synchrotron component. If so,
spinning dust, in conjunction with a synchrotron component, could
explain the cm-excess and alleviate the need for a synchrotron
break.


%

\subsection{Extinction due to mm-sized metallic needles?} \label{sec:needles}

Alternatively the 250~GHz deficit could be interpreted as extinction
due to dust. For standard dust grains such millimetre extinction would
require dust-to-gas mass ratios of order 1. However, metallic needles
bypass the mass problem for extinction, and explain the 250~GHz
deficit with gray extinction at wavelengths shorter than 1~mm, and an
opacity decreasing as $\lambda^{-2}$ towards 1~cm.  Using the formulae
given in \citet{dwe04a, dwe04b}, a long-wavelength cutoff for the gray
extinction of $\lambda_\circ = 1~$mm requires a needle aspect ratio
$l/a$ of $\sim$8000 for a resistivity of $\rho_R =
10^{-6}~\Omega$~cm. Imposing a unit needle opacity at 1~mm gives a
total needle mass of $2\,10^{-6}$~M$_{\odot}$ for a grain material
density of $\rho_m = 8.15~$g~cm$^{-3}$, i.e. much less than typical
nebular masses of $\sim 0.1~$~M$_{\odot}$. Yet, to be seen in
absorption at 1~mm the needles would have to be extremely cold, about
1~K. Thermal balance gives lower temperatures for longer needles (Eli
Dwek, private communication), and although needles are colder than
normal interstellar grains \citep[the needle temperature would be 8~K
in Cas A,][]{dwe04a}, temperatures $<1~$K seem unphysical. A solution
may lie in modelling the needle extinction in a distribution of needle
aspect ratios, such that there exist a very small population of
ultra-cold grains.

We attempted to fit the 31--250~GHz drop in NGC~6369 with needle
extinction. The needle extinction is applied as a screen to the
emergent flux.  For simplicity we added the needle emission as if it
were a foreground to the free-free nebula. A fit to the data is
obtained with $\lambda_\circ = 1.2~$mm, maximum opacity 0.82, and $T =
1.2~$K. The free-free component corresponds to a uniform slab with
$N_e = 3373 \pm 85$~cm$^{-3}$, $T_e = 6580\pm1031~$K.

The SEST 99~GHz and 147~GHz heterodyne data shown on
Fig.~\ref{fig:NGC6369}c are in very good agreement with needle
extinction. They were originally left out from the fit, and added at a
later stage for comparison purposes. The SEST heterodyne data are
subject to numerous uncertainties. The possibility of needle
extinction will not be considered further in this work because it
requires extreme temperatures.

\section{Comments on individual objects} \label{sec:indiv}

In this Section we fit the SEDs of selected PNe with analytic
radiative transfer solutions.  All fits (except the single component
free-free spectra) include modified black bodies constrained to the
60- and 100-$\mu$m {\em IRAS} flux densities, with a dust emissivity
index of $\beta = 2.0$. This choice for $\beta$ may seem rather high
compared to the canonical ISM value of $\beta = 1.7$. Yet higher
emissivities minimise the contribution of the Rayleigh-Jeans tail to
the 250~GHz flux density, thereby minimising the 250~GHz deficit
relative to the free-free level expected from 31~GHz.  The PNe listed
in Table~\ref{table:master} were selected on their being
IR-bright. Their flux densities peak in the {\em IRAS}~25--{\em
  IRAS}~60 bands, and decrease from 60 to 100$\mu$m.

Unless otherwise stated the low frequency measurements are extracted
from \citet[][ at 5 and 14~GHz]{mil82}, \citet[][ at 408~MHz]{cal82},
and \citet[][ at 1.4~GHz]{con98}. We generally preferred single-dish
measurements, when available, in the cases where the interferometric
data are resolved.  PN distances are notoriously difficult to measure,
and are usually assigned 100\% uncertainties.

\subsection{The largest 250~GHz deficits}


\subsubsection{NGC~7009}

Few Galactic nebulae can be found near NGC~7009, at 21~h. Its CBI and
SIMBA flux densities are the best measurements reported in this
work. The 250~GHz value is averaged from 3 different observing
runs. The 31~GHz value is averaged over 7 different nights and spread
over 3 years.

\begin{figure}
\begin{center}
\includegraphics[width=\columnwidth,height=!]{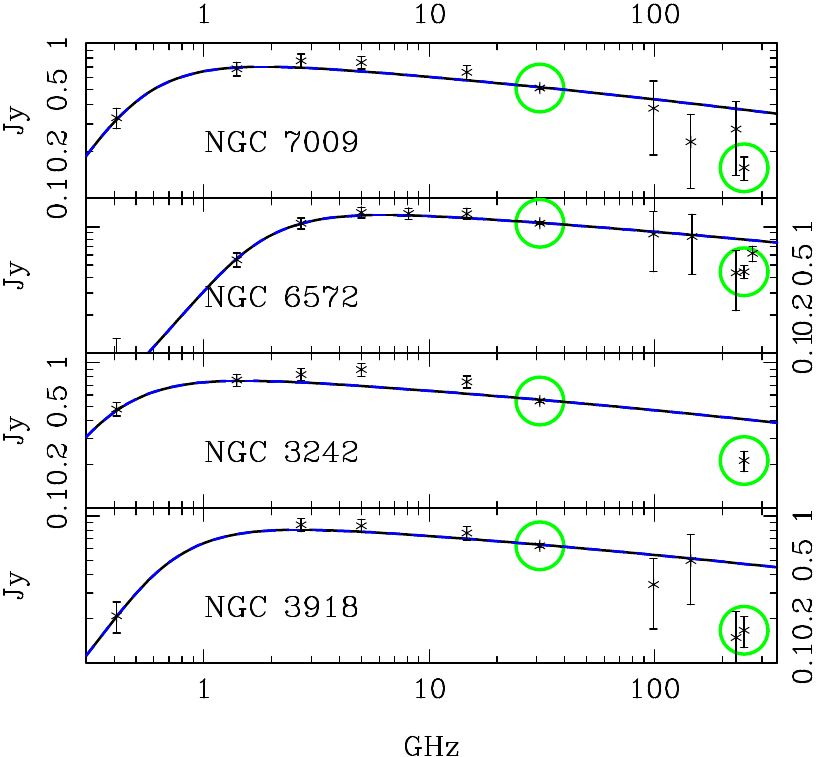}
\end{center}
\caption{\label{fig:deficits} SEDs of the PNe with the largest 250~GHz
  deficits. We plot flux density in Jy ($y-$axis) as a function of
  frequency in GHz ($x-$axis). The CBI and SIMBA datapoints are
  indicated by circles. The solid line is a single uniform-slab
  free-free component, drawn to emphasize deviations from free-free
  spectra, and fit to the $\le 31~$GHz data with the following
  parameters: {\bf NGC~7009}, $T_e = 4550\pm943$~K, $N_e =
  2618\pm81$~cm$^{-3}$; {\bf NGC~6572}, $T_e= 8589\pm1278~$K, $N_e =
  14810\pm350$~cm$^{-3}$; {\bf NGC~3242}, $T_e = 7921 \pm 1601~$K,
  $N_e = 2938\pm92~$cm$^{-3}$; {\bf NGC~3918}, $T_e = 9750\pm2447$~K,
  $N_e = 4501\pm185~$cm$^{-3}$.}
\end{figure}

The SED of NGC~7009 in Fig.~\ref{fig:deficits} is very similar to
NGC~6369. As in Sec.~\ref{sec:sync}, a fit to the data can be obtained
with a broken synchrotron spectrum, with a low frequency index of
$\alpha_\mathrm{LF} = -0.4$, and a high frequency index $\alpha =
-1.0$ absorbed by a free-free screen with $\langle
N_e(\mathrm{screen}) \rangle = 345\pm54$~cm$^{-3}$,
$T_e(\mathrm{screen}) = 50~$K, and a power-law opacity distribution
with $\gamma = 6.0$. The core free-free component would be $N_e =
1456\pm86$~cm$^{-3}$, $T_e = 10^{4}$ ~K. We used a distance of 420~pc,
and a nebular diameter of 28.5~arcsec \citep[as compiled by][]{ack92}.

\subsubsection{NGC~6572}

The SED of NGC~6572 in Fig.~\ref{fig:deficits} is again very similar to
NGC~6369. We adopt the expansion parallax distance of
$1.49\pm0.62$~kpc \citep{haj95}, and a nebular diameter of 8~arcsec
\citep[as observed with the VLA by][]{zij89}.

We could fit the data with an attenuated and broken synchrotron
spectrum, with a low-frequency spectral index of $-0.4$, and a high
frequency index of -1.0.  The bulk free-free component in
Fig.~\ref{fig:deficits} has $N_e = 10056\pm162$~cm$^{-3}$ and $T_e =
10^{4}$~K. We used a free-free screen at 50~K, with
$\langle N_e(\mathrm{screen}) \rangle = 998\pm248$, and power-law opacity distribution
with index $\gamma = 5.2$. The sub-mm dust has a temperature of 76~K.

The data from \citet{hoa92} confirm our calibration of the SIMBA flux
densities. In can be seen in Fig.~\ref{fig:deficits} that the SIMBA
data point, at 250~GHz, is low but agrees with the \citet{hoa92} data
within the uncerainties. The rising sequence with frequency could
perhaps be due to the Rayleigh-Jeans tail of sub-mm dust extending to
250~GHz. We have also plotted on Fig.~\ref{fig:deficits} data points from
the preliminary SEST measurements (assigned exagerated error bars).

\subsubsection{NGC~3242}

NGC~3242 has an expansion parallax distance of $420\pm160$~pc
\citep{haj95}, and a nebular diameter of about 25~arcsec. We could fit
the SED of NGC~3242 with the same synchrotron component as for
NGC~6369, except the core free-free emission is constrained to the
250~GHz flux density, has $N_e = 2401$~cm$^{-3}$, and is assigned a
temperature of 20\,000~K. A free-free screen at 20~K has an average
density of 200~cm$^{-3}$, and a power-law distribution of opacities
with $\gamma = 10$. A single power law fit at 5--31~GHz gives a
exponent of $-0.290^{-0.237}_{-0.340}\pm0.051$, which is different
  from the free-free $\alpha_{5}^{31}$ value by more than 3~$\sigma$.

\subsubsection{NGC~3918}

%
%

The high excitation PN NGC~3918 has been studied in depth by
\citet{cle87}. We adopt their preferred distance of 1.5~kpc, and a
nebular angular size of 15~arcsec. The SED of NGC~3918 show on
Fig.~\ref{fig:NGC6369} can be fit by a broken synchrotron spectrum,
with spectral indices of $-0.4$ and $-1.0$. The free-free screen is at
50~K, and has an average density of $\langle N_e(\mathrm{screen})
\rangle = 371 \pm 91$~cm$^{-3}$. The bulk of the 250~GHz flux density is due to
the traditional free-free emission, at 10$^{4}$~K and $N_e =
2342\pm306$~cm$^{-3}$.

%
%
%
\subsection{Outliers}

In this section we discuss the objects that we have considered as
outliers from the correlations described in Sec.~\ref{sec:comp}. They
differ in physical characteristics and radio spectra from the rest of
the sample.

\subsubsection{M~2-9} \label{sec:M2-9}

M~2-9 is a peculiar object. Its ionizing central star is undergoing
sustained mass loss, as revealed by broad H$\alpha$ wings
\citep[e.g.][]{arr03,swi79}. As reported by \citet{kwo85}, the radio
SED of M~2-9 is that of an optically thick stellar wind, and is in
marked contrast with the other objects studied here. 

\begin{figure}
\begin{center}
\includegraphics[width=\columnwidth,height=!]{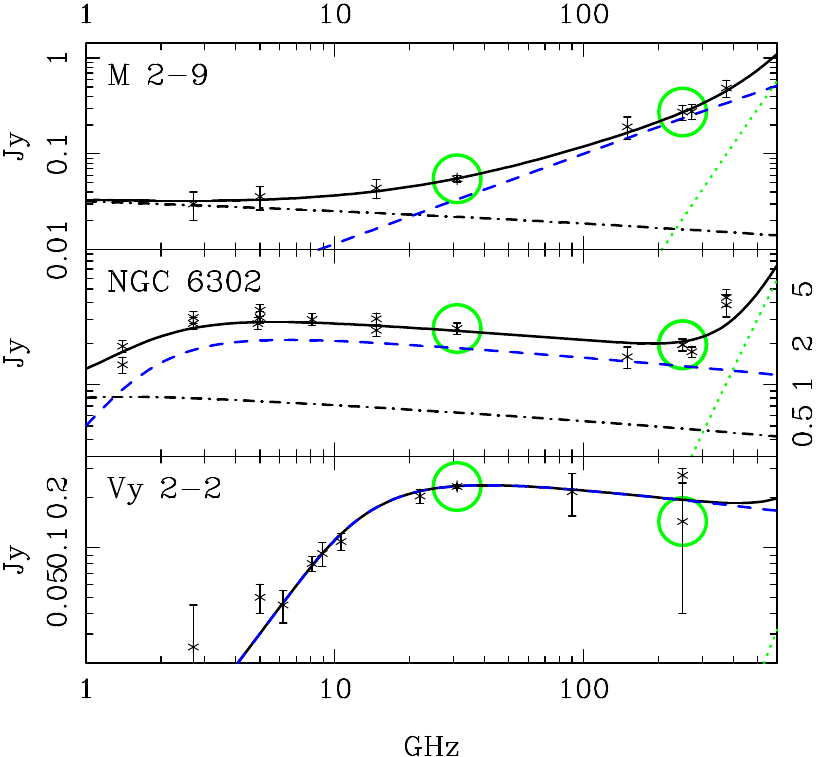}
\end{center}
\caption{\label{fig:outliers} SEDs of outliers. We plot flux density
  in Jy ($y-$axis) as a function of frequency in GHz ($x-$axis). The
  CBI and SIMBA datapoints are indicated by circles. {\bf M\,2-9} The
  components that sum up the model spectrum (in solid line) are a
  modified black body (dotted line), an optically thick stellar wind
  (dashed line) and a uniform slab (dash-dotted). {\bf NGC~6302} We
  use a core-halo model to fit the observed spectrum. The core
  component is shown in dashed line ($N_e = 13470^{-3}$, $T_e =
  8917$~K), and the halo component is shown in dash-dotted line ($N_e
  = 4270$~cm$^{-3}$, $T_e = 20000~$K, with a diameter of
  20~arcsec). This model is not unique and only serves the purpose of
  indicating a physical reference. {\bf Vy 2-2} Line styles follow
  from the NGC 6302 spectrum. The flux densities are extracted from
  \citet{pur82} and \citet{alt94}.}
\end{figure}

In Fig.~\ref{fig:outliers} we model the radio spectrum of M~2-9 with a
power-law distribution of opacities, following \citet{cal91}, except
the opacity law we used is $\tau(\omega) = \delta \omega^{-\gamma}
$. As explained in \citet{cal91}, the detailed spatial distribution of
opacities is irrelevant to the observed free-free flux density. Only
the distribution of opacities as a function of a single solid angle
variable is required. In general the opacity will be a function of
some 1-D parametrisation of solid angle, $\omega$. As an example
definition of $\omega$, \cite{cal91} pixelates the sky emission of the
nebulae and orders pixels by their opacities. A natural choice for
$\omega$ in spherically symmetric nebulae is the nebular radius.  In
the case of the optically thick stellar wind in M~2-9, the bulk of the
emission at cm- to mm-wavelengths comes from the wind, with a power
law profile: $\tau(\omega) = \delta \omega^{-\gamma}$. A clumpy
free-free screen can also be modelled with a similar opacity profile
as M~2-9, except the clumps are scattered across the nebula. Larger
values of $\gamma$ correspond to smaller fractions of the nebular
pixels containing most of the opacity. Power law profiles allow for a
smoother transition to the optically thick frequency regime than
uniform slabs \citep[a proof is given in][]{cal91}.

The emergent flux density is
\begin{equation}
F_\nu = \Omega_N B_\nu \int_{\omega_u}^1 d\omega (1-\exp(-\tau_{\nu}(\omega))),
\end{equation}
where $\Omega_N$ is the nebular solid angle, $\omega = \Omega /
\Omega_N $ is a solid angle variable (for instance angular distance
from the central star), $\omega_u \ll 1$ is fixed by requiring
$\int_{\omega_u}^1 d\omega \tau(\omega) = 1$, and
\begin{equation}
\tau_\nu(\omega) = \langle \tau_\nu \rangle \tau(w). 
\end{equation}

We compiled the SED of M~2-9 with the data in \citet{hoa92}, our own
CBI and SEST measurements, the 5~GHz and 14~GHz data from
\citet{mil82} and the 2.7~GHz flux density from \citet{cal82}.  We
obtain a fit to the radio-IR spectrum of M2-9 with a superposition of
a uniform slab nebula to account for the flattening at low
frequencies, a power law distribution with $\delta = 1.6~10^{-9} $,
$\gamma = 2.06 $ to model the optically thick core, an average core
density of $\langle Ne \rangle = 2093$~cm$^{-3}$, and a temperature
$T_e = 7700~$K. The best fit average emission measure entering in the
average opacity $\langle \tau_\nu \rangle$ is $\langle \mathrm{EM}
\rangle = 1.3~10^6$ pc~cm$^{-6}$, with a total nebular diameter of
1~arcmin. We take the agreement between the SIMBA flux density of
M\,2-9 with the data from \citet{hoa92} and \citet{alt94}, as well as
with the model SED, as a confirmation of our data calibration and
reduction procedure.


\subsubsection{NGC~6302} \label{Sec:NGC6302}

NGC~6302 is the highest excitation PN known, and is probably derived
from the upper mass range of planetary nebula progenitors
\citep[e.g.,][]{cas00}.  As such it has been the focus of much
study. The ionised nebula is surrounded by copious amounts of neutral
and molecular material \citep{rod82}. 

In Fig.~\ref{fig:outliers} we summarise the available dataset on
NGC~6302. The flux densities have been extracted from
\citet{hoa92,rod85, ter74, con98, zij89, mil82, mil79, cal82}. We
adopt the expansion parallax distance of $1.6\pm0.6$~kpc
\citep{gom93}, and a core nebular diameter of 10~arcsec \citep{zij89},
surrounded by a halo extending up to $\sim~1$arcmin. As in the case of
M\,2-9, we find that the SIMBA data point agrees, within the error
bars, with the data from \citet{hoa92}.

We fit the SED of NGC~6302 with a core-halo model, as in
\citet{rod85}, including the dust temperature as a variable. The {\em
IRAS}~100- and 60-$\mu$m data alone give $T_\mathrm{dust}=46~$K, if
the dust emissivity is $\beta = +2$. Including the UKIRT and JCMT data
from \citet{hoa92}, as well as our SIMBA point, gives a dust
temperature of 42~K. We believe there is room for a colder dust
component, that could be present at $\sim$200~GHz. 

The same situation occurs in NGC~6537, often considered as a slightly
lower excitation twin of NGC~6302. The Rayleigh-Jeans tail of the
sub-millimetre dust could contribute to the 250~GHz flux.

NGC~6302 is an important cross-check on the SIMBA calibration
scale. Raising the SIMBA flux densities by 50\% would eliminate the
cm-excess, and result in a mm-excess in some objects. But the SIMBA
data on NGC~6302 is consistent with the literature. Raising the
NGC~6302 250~GHz flux densities by 50\% could be inconsistent with
\citet{hoa92} by 5~$\sigma$.

%

%
%
\subsubsection{Vy~2-2}


Vy~2-2 is a very young low excitation PN \citep[e.g.,][]{liu01}, with
strong 10$\mu$m silicate emission at $\sim$365~K \citep{ait82}.
Faint, compact and opaque radio emission at the nucleus of Vy~2-2 is
resolved into a 0.4~arcsec diameter ring at 22~GHz \citep{chr98}, and
is surrounded by a faint 25~arcsec H$\alpha$ halo.

Its distance is greater than 1~kpc \citep{chr98}, and traditional
indicators place it at 7.9~kpc \citep{zha95} . Since Vy~2-2 is a young
object, we adopt the distance of 3.9~kpc, given by the {\em IRAS}
fluxes and the tip-of-the-AGB luminosity function \citep{cas01}.

Vy~2-2 is very IR bright, while optically thick and faint at radio
frequencies. The large uncertainty on its {\em SIMBA} flux density
left it out of Fig.~\ref{fig:fracexcess}. Its {\em IRAS~60~$\mu$m} to
5~GHz ratio is 0.86~10$^{-3}$, 3-4 times higher than even the largest
values in Fig.~\ref{fig:fracexcess} or in NGC~6302.

The radio SED of Vy~2-2, shown on Fig.~\ref{fig:outliers} can be fit
with a simple free-free model consisting of a uniform slab, with $T_e
= 8522\pm 660$, $N_e = 3~10^{5}\pm4~10^{3}~$cm$^{-3}$, and a depth of
0.01~pc. It can be appreciated this SED is very different from the
rest of the PNe considered in this work.

%



\subsubsection{NGC~2440}

NGC~2440 is more evolved than the rest of our sample, and its mid-IR
flux is due to ionic lines. Dust in this object radiates only in the
far-IR, and is presumably located further from the central star than
in the other PNe. The SIMBA data point is unfortunately very
uncertain, and so is the $<31~$GHz data. We mention it here because
the low-frequency data seems to peak at 5~GHz and drop with frequency
steeper than allowed by free-free emission at low-frequencies, in a
fashion that is reminiscent of the 250~GHz deficit evident in
NGC~6369.

\section{Conclusions} \label{sec:conc}

We have found an inconsistency with the free-free paradigm for the
radio continua of prototypical PNe, in the form of an excess over
free-free emission at 31~GHz when compared to the 250~GHz flux
densities. The fraction of the 31~GHz continuum not due to free-free
emission, average over the 10 objects with less than 30\%
uncertainties, is $51\pm3$\%, with a weighted scatter of 11\%. The
cm-excess is present in all objects that are optically thin at
31~GHz, with the exception of NGC~6302 and NGC~6537, whose sub-mm dust
tails reach 250~GHz. The most significant detections are in NGC~6369,
NGC~7009, NGC~3242 and NGC~6572. 

The evidence for the cm-excess stems mostly from the 250~GHz data
acquired with SIMBA. Raising the SIMBA calibration scale by $\sim$50\%
would bring some of the 250~GHz levels in agreement with free-free,
but would also render the data on NGC~6302 and M~2-9 incompatible with
the literature. Additional evidence can be found in the 5--31~GHz
spectrum of NGC~3242, and in the average $\langle \alpha_{14}^{31}
\rangle$, which are steeper than free-free.

The 31~GHz excess follows a loose correlation with the {\em IRAS} flux
densities.  Yet the examination of individual objects shows that
spinning dust cannot alone explain the low frequency data, at least
with the models currently available. Needle extinction at 250~GHz
would require extreme temperatures. A synchrotron component is a
possibility, if it is attenuated at low-frequencies by a cold
free-free screen ($<500~$K). The cm-wave - infrared correlation could
be due to synchrotron radiation originating in compact and dusty
inclusions exposed to the fast stellar winds observed in PNe.

\section*{Acknowledgments}

We are grateful to Pat Roche for useful discussions and for reminding
S.C. about Helium, to Eli Dwek for an email exchange providing
extensive information on his needle models, and to Wouter Vlemmings
for feedback on the nebular magnetic fields. The Swedish-ESO
Submillimetre Telescope, SEST, was operated jointly by ESO and the
Swedish National Facility for Radioastronomy, Onsala Space Observatory
at Chalmers University of Technology. S.C. acknowledges support from
FONDECYT grant 1060827, and from the Chilean Center for Astrophysics
FONDAP 15010003.  We gratefully acknowledge the generous support of
Maxine and Ronald Linde, Cecil and Sally Drinkward, Barbara and
Stanely Rawn, Jr., Fred Kavli, and Rochus Vogt.  Part of the research
described in this paper was carried out at the Jet Propulsion
Laboratory, California Institute of Technology, under a contract with
the National Aeronautics and Space Administration. CD thanks B.  and
S. Rawn Jr. for funding a fellowship at the California Institute of
Technology for part of this work.

\appendix

\section{Thermal spectral indices} \label{sec:exactindex}

The free-free emissivity $\epsilon$ for electron-ion encounters for a
dilute plasma \citep[at frequencies much larger than the plasma
frequency,][]{sch60} can be written as \citep[e.g.][per unit solid
angle]{bec00},
\begin{equation}
\epsilon = N_e N_i \frac{8}{3} \sqrt{\frac{2 \pi}{3}} \frac{e^6
  Z_i^2}{(m c^2)^{(3/2)}} \frac{1}{\sqrt{kT}} \exp\left(-\frac{h\nu}{k
  T}\right) g_\mathrm{ff}, \label{eq:ffemis}
\end{equation}
where $N_i$ is the ion density and $g_\mathrm{ff}(\nu,T_e,Z_i)$ is a
frequency- and temperature-dependent Gaunt factor. 

\subsection{Exact Gaunt factors for free-free emission.}

We searched the literature for approximations to the exact
$g_\mathrm{ff}(Z_i,\nu)$, and found errors comparable to the level of
precision in our spectral analysis (of $\sim$5--10\%). We thus chose
to calculate exact Gaunt factors, using the formulae and procedures
described in \citet{gra58}, which have also been reproduced in
\citet{bec00} and \citet{mun06}.

The form we used for the non-relativistic Gaunt factor in encounters
between electrons and bare nuclei is that given by
\citet{men35}. Velocity-dependent Gaunt factors for individual
encounters involve the evaluation of complex hypergeometric functions
$_2F_1(a,b,c,z)$. We used the hypergeometric series when possible
\citep[for $|z| < \mathrm{min}(0.1, |0.1 c / (ab) |$ , as prescribed
by][]{bec00}, and integrated the hypergeometric equation otherwise. We
used the GNU Scientific Library routines for ODEs, finding that the
Bulirsch-Stoer method gave best results. We used the prescription in
\citet{bec00} as starting point for the ODE integration.  For $|z| >
1$ we used the `linear transformation formulae' from Section~15.3 of
\citet{abr64} to keep the complex argument $z$ within the unit
circle. 

Our hypergeometric routines are robust for any frequency for
relatively fast electrons, when the parameter $\eta_i = -\alpha Z (c /
v_i) < 1$, in the notation of \citet{gra58}. But the ODE integration
crashed for large $a$ and $b$, corresponding to slow incident
electrons, or when the parameter $\eta_i > 100$. So we tried the
hypergeometric routines in \citep[][NR]{pre96}, edited to double
precision and with the starting point of \citet{bec00}, finding the
same numerical limitation as for GSL.

The domain of slow incident electrons corresponds to the so-called
``classical'' limit, when the de Broglie wavelength is small compared
to $Z\,e^2/(m\,v_i^2)$, a measure of size of the Coulomb field
\citep[e.g.][]{bru62}. In what follows we will refer to the ``classical''
limit as the WKB limit, and use the term classical for the classical
mechanics calculation leading to the Gaunt factors used by
\citet{sch60} and \citet{ost61}. The WKB limit involves Hankel
functions with imaginary arguments, and are rather difficult to
evaluate.

\begin{figure}
\begin{center}
\includegraphics[width=0.9\columnwidth,height=!]{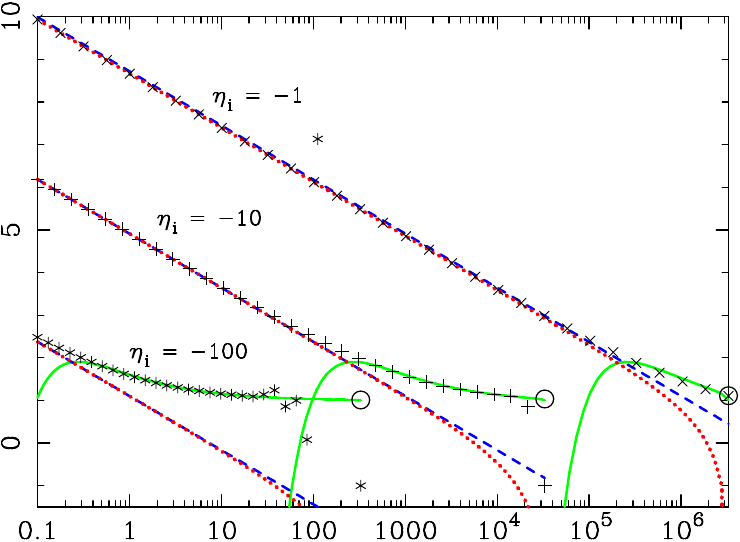}
\end{center}
\caption{\label{fig:gsamp} The free-free Gaunt factor for individual
  electron-proton encounters with incident velocities corresponding to
  $\eta_i = -1,-10$ and $-100$. We plot the exact calculation (with
  'x','+','*' symbols, respectively), the LFL (dotted line), the
  classical (dashed line), and WKB Gaunt factors (solid line). The
  recombination limit Gaunt factor \citep[Eq.~6 from][]{bru62} is
  indicted by a circle.}
\end{figure}

Fortunately, a good and simple approximation for $\eta_i > 100$ can be
obtained in the following way. When the product photon has a frequency
close to the recombination (bound-free) limit, the WKB Gaunt factor
reduces to a very simple expression given by Eq.~14 from
\citet{gra58}. In the limit of low-frequencies, the incident electron
follows a classical trajectory unperturbed by the emission of the
radio photon, and the Oster/Scheuer Gaunt factors are applicable. At
frequencies between 0.1 and 1000~GHz $\eta_i$ varies between 0.3 and
2500 when $T_e = 1-50000~$K.  Fig.~\ref{fig:gsamp} shows the free-free
Gaunt factor for individual encounters involving incident electrons
with $\eta_i$ of $-1,-10$, and $-$100, as a function of
frequency. Such values of $v_i$ correspond to the rms velocities at
temperatures of $\sim 10^5,10^3,10$~K, respectively. The maximum
frequency corresponds to the recombination limit $\nu_g$. We compare
the exact calculation, in asterisks, with the low-frequency limit
\citep[hereafter LFL, as given in,][their Eq.~23, but replacing the
absolute value by a minus sign]{bec00}\footnote{The LFL was originally
derived by Elwert, as explained by \citet{ost61}}, the
WKB and the classical Gaunt factor. \citet{ost61} shows that the LFL
converges to the classical limit for large $\eta_i$.  It can be seen
that as $\nu$ approaches $\nu_g$ the exact calculation scatters about
the expected WKB value, especially for large $\eta_i$. This is because
the ODE integration becomes numerically unstable, and its results
cannot be trusted, as $\nu$ approaches $\nu_g$. The ODE integration
altogether crashes for even larger values of $\eta_i$.  But there is
always a regime where an approximation can be substituted for the
exact calculation, and this value is the largest between the classical
and WKB Gaunt factors. Thus, we used the WKB Gaunt factor when $\nu >
0.01\nu_g$ and $\eta_i < -1$, and the maximum between the classical
and WKB Gaunt factors when $\eta_i < -100$. This is different from the
choice of \citet{bec00}, who always used the LFL instead of the WKB
approximation. The relative difference between the use of LFL or WKB
is important only at very low temperatures, and is within $\sim
10^{-3}$ when $T_e > 1000~$K.

The Maxwellian integration for the total rate coefficients (a.k.a. the
average coefficients), Eq.~24 in \citet{gra58}, was carried out using
the PDL interface to the GNU Scientific Library. In order to confirm
our calculation we did the exercise of using the LFL of \citet{bec00}
for large $|\eta_i|$ (i.e. not the WKB limit), and found that we
obtain the same total Gaunt factors within a relative error of $\sim
10^{-5}$ as the calculations in \citet{mun06}, who used the {\em
Mathematica} package to directly evaluate the \citet{men35} Gaunt
factors \citep[and following][]{bec00}.

\begin{figure}
\begin{center}
\includegraphics[width=0.6\columnwidth,height=!]{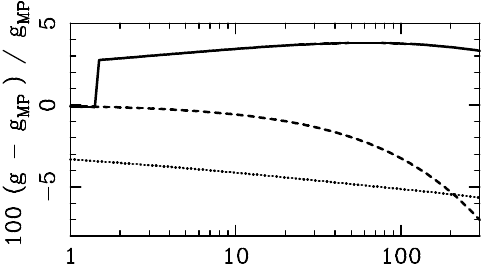}
\end{center}
\caption{Comparison between various gaunt factor formulae at $T_e =
  7000~$K.  We plot the percentual difference between the Gaunt
  factors $g$ for $Z=2$ (dotted), the \citet{ost61} (dashed) and
  Cloudy (solid) \citep{fer98} approximations, and the exact value
  with $Z=1$ from \citet{men35}, $g_\mathrm{MP}$.  For reference we
  obtain $g_\mathrm{MP}(\nu = 250~\mathrm{GHz}, T_e = 7000~\mathrm{K},
  Z = 1) = 2.7954$.  The curve for $Z=2$ has been divided by
  2.\label{fig:compargff}}
\end{figure}

The use of common approximations to the radio-frequency Gaunt factors
involve errors $\sim 5$\% in the 1-300~GHz frequency range for
traditional nebular temperatures. In Fig.~\ref{fig:compargff} we have
compared the exact Gaunt factors with the approximations in
\citet{sch60,ost61}, and the Cloudy photoionisation package
\citep[version c06.02c, last described by][]{fer98}, which uses
free-free Gaunt factors based on \citet{hum88}. Note that the
\citet{alt60} free-free opacities, with the $a(\nu,T)$ factor
correction from \citet{mez67}, are equal to free-free opacities
calculated directly from the \citet{ost61} Gaunt factors.

\begin{figure}
\begin{center}
\includegraphics[width=0.8\columnwidth,height=!]{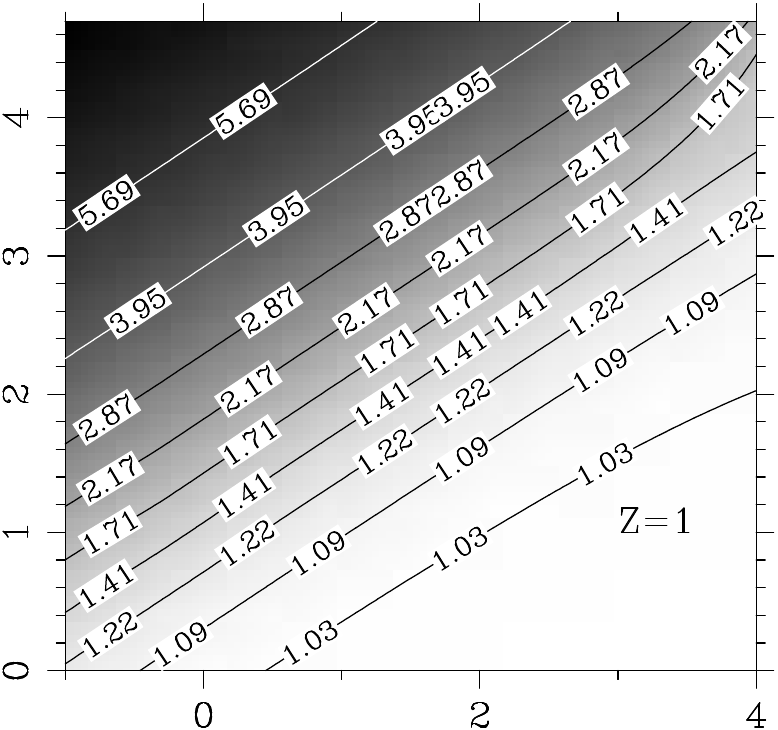}
\end{center}
\caption{Exact (non-relativistic) Gaunt factors for $Z=1$, as a
  function of $\log(\nu/\mathrm{GHz})$ in $x-$axis and
  $\log(T_e/\mathrm{K})$ in $y-$axis \label{fig:grids}}
\end{figure}

For practical applications, in order to speed up the optimisations
involved in fitting the PN SEDs, we gridded the exact free-free Gaunt
factors in the $(\log(T_e),\log(\nu))$ space. Fig.~\ref{fig:grids}
shows the corresponding Gaunt factor images. It can be appreciated
that there are very little differences between the $Z=2$ and $Z=1$
Gaunt factors. The limit for a cold plasma, at high frequencies, is 1,
as expected. Note that the Gaunt factors given by \citet{ost70} tend
to 0 for cold plasmas because they include the $\exp(-h\nu/(kT))$
term, which is usually associated with the emissivity
(Eq.~\ref{eq:ffemis}) rather than with the Gaunt factor.

\subsection{Free-bound emission}

Since we are considering thermal emission from cold plasma, the
possibility of radio free-bound emission must be taken into
account\footnote{we have not included the contribution of radio
  recombination lines (RRLs) to broad band data. RRLs could be
  important at $\sim$300~GHz for $T_e \sim 20~$K}.  We use the
formulae compiled in \citet{bru62}, taking free-bound Gaunt factors of
1. Fig.~8 of \citet{bru62} shows that, in the radio-sub-mm regime, the
free-bound Gaunt factors are essentially 1 for recombinations to any
principal quantum numbers.

\subsection{Contribution from He and representative  PN spectral indices.}

\label{sec:finalindices}

\begin{figure}
\begin{center}
\includegraphics[width=0.9\columnwidth,height=!]{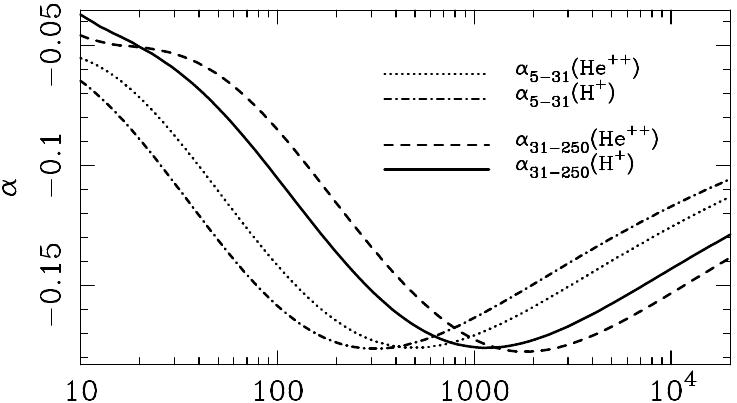}
\end{center}
\caption{Exact (non-relativistic) optically-thin thermal spectral
  indices for encounters between electrons and bare H and He nuclei,
  evaluated for a range of temperatures, and two pairs of frequencies:
  31-250~GHz, and 5-31~GHz.  \label{fig:ffindex}}
\end{figure}

The free-free and free-bound emissivities depend on temperature. In
Figs.~\ref{fig:ffindex} we summarise the variations of
$\alpha_5^{31}(\mathrm{H}^+)$, $\alpha_5^{31}(\mathrm{He}^{++})$,
$\alpha_{31}^{250}(\mathrm{H}^+)$, and
$\alpha_{31}^{250}(\mathrm{He}^{++})$, in the case of optically-thin
emission and for a range of electron temperatures $T_e$.

The Gaunt factors for the He free-free continuum should be almost
identical to the H values in the domain where the wavelength of the
electron incident on He$^{+}$ has a wavelength much larger than the
He$^+$ Bohr radius. Thus, we can neglect differences with He up to a
temperature $Te = (2 h / a_\circ)^2 / (3 k \, m_e) \approx
17~10^6$~K. As explained above, the Gaunt factors for the He$^+$
free-free continuum are slightly different from H. Considering that
the most He rich nebulae (of Peimbert type I) reach He abundance by
number of 18\% relative to H, a fraction of which is doubly ionised,
we can conservatively assume a contribution from the He$^+$ continuum
of up to 20\%, and a minimum of zero.  This translates into the
spectral index uncertainties given in Table~\ref{table:indices}, where
we have assumed an average abundance of He$^{++}$ of 10\% relative to
H$^+$.

\label{lastpage}

\end{document}